\documentclass[%
reprint,
superscriptaddress,
 aps,
 pra,
longbibliography,
dvipsnames,
twocolumn,
xcolor=table
]{revtex4-1}
\usepackage[main=english]{babel}

\usepackage{dcolumn}% Align table columns on decimal point
\usepackage{bm}% bold math
\usepackage{siunitx}
\usepackage{dsfont}
\usepackage{caption}
\usepackage{braket}
\usepackage{amsthm}
\usepackage{mathtools}
\usepackage{physics}
\usepackage{comment}
\usepackage{multirow}
\usepackage{algpseudocode}
\usepackage{amsthm}
\usepackage{algorithm}

\algnewcommand{\Inputs}[1]{%
  \State \textbf{Inputs:}
  \Statex \hspace*{\algorithmicindent}\parbox[t]{.8\linewidth}{\raggedright #1}
}
\algnewcommand{\Outputs}[1]{%
  \State \textbf{Outputs:}
  \Statex \hspace*{\algorithmicindent}\parbox[t]{.8\linewidth}{\raggedright #1}
}
\algnewcommand{\Initialize}[1]{%
  \State \textbf{Initialize:}
  \Statex \hspace*{\algorithmicindent}\parbox[t]{.8\linewidth}{\raggedright #1}
}

\usepackage{tikz}
\usetikzlibrary{quantikz}

\usepackage{mathtools}
\usepackage{ragged2e}

\usepackage[utf8]{inputenc}
\usepackage[toc,page]{appendix}

\usepackage{graphicx}
\usepackage{hyperref}
\hypersetup{colorlinks}
\usepackage{soul}

\usepackage{bbold}
\usepackage{amssymb}
\usepackage{float}
\usepackage{hyphenat}

\newcommand{\update}[1]{{ #1}}

\begin{document}

\newcommand{\orcidicon}[1]{\href{https://orcid.org/#1}{\includegraphics[height=\fontcharht\font`\B]{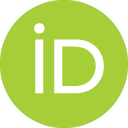}}}

\author{Paulin de Schoulepnikoff}
\affiliation{Institute of Physics, Ecole Polytechnique Fédérale de Lausanne (EPFL), CH-1015 Lausanne, Switzerland}
\affiliation{European Organization for Nuclear Research (CERN), Geneva 1211, Switzerland}

\author{Oriel~Kiss\orcidicon{0000-0001-7461-3342}}
\email{oriel.kiss@cern.ch}
\affiliation{European Organization for Nuclear Research (CERN), Geneva 1211, Switzerland}
\affiliation{Department of Nuclear and Particle Physics, University of Geneva, Geneva 1211, Switzerland}

\author{Sofia Vallecorsa\orcidicon{0000-0002-7003-5765}}
\affiliation{European Organization for Nuclear Research (CERN), Geneva 1211, Switzerland}

\author{Giuseppe Carleo}
\affiliation{Institute of Physics, Ecole Polytechnique Fédérale de Lausanne (EPFL), CH-1015 Lausanne, Switzerland}

\author{Michele~Grossi\orcidicon{0000-0003-1718-1314}}
\email{michele.grossi@cern.ch}
\affiliation{European Organization for Nuclear Research (CERN), Geneva 1211, Switzerland}

%\title{Learning Schmidt decompositions with artificial neural networks and quantum circuits} 

% \title{A hybrid approach for learning Schmidt decompositions}
\title{Hybrid Ground-State Quantum Algorithms based on Neural Schrödinger Forging}

\date{\today}

\begin{abstract} 

Entanglement forging based variational algorithms leverage the bi-partition of quantum systems for addressing ground state problems. The primary limitation of these approaches lies in the exponential summation required over the numerous potential basis states, or bitstrings, when performing the Schmidt decomposition of the whole system. To overcome this challenge, we propose a method for entanglement forging employing generative neural networks to identify the most pertinent bitstrings, eliminating the need for the exponential sum. Through empirical demonstrations on systems of increasing complexity, we show that the proposed algorithm achieves comparable or superior performance compared to the existing standard implementation of entanglement forging. Moreover, by controlling the amount of required resources, this scheme can be applied to larger, as well as non\hyp permutation\hyp invariant systems, where the latter constraint is associated with the Heisenberg forging procedure. We substantiate our findings through numerical simulations conducted on spin models exhibiting one-dimensional ring, two-dimensional triangular lattice topologies, and nuclear shell model configurations.

\end{abstract}

\maketitle

\section{Introduction}

In recent years, significant advances have been made in simulating the static and dynamical properties of many\hyp body quantum systems using variational algorithms. For instance, \update{density matrix renormalisation group} methods based on matrix\hyp product states \cite{MPS_Cirac_2008,MPS_Dukelsky,MPS_Rommer}, neural networks quantum states \cite{Carleo_Science}, equivariant neural networks \cite{FermiNet} or kernels methods \cite{carleo_kernel} have accurately computed the ground state energy of spin systems \cite{NNQS_magnets}, or also fermi systems, such as molecules \cite{carleo_chemistry_nnqs} or nuclei \cite{Lovato_nuclear_nnqs}. While neural network quantum states represent wave functions using classical representations, we can also consider their quantum counterpart, where the wave function ansatz takes the form of a parametrized quantum circuit \cite{variational_quantum_algorithm}, such as in the popular variational quantum eigensolver (VQE) \cite{vqe_original}. Even if VQE has been successfully applied in various areas, such as chemistry \cite{vqe_original,VQE_Gambetta,UCC-chemistry}, spin chains \cite{VQE_magnet,magnetogrossi,Monaco_PRB,LMG} or nuclei \cite{Papenbrock-Deuterium,Be8-VQE,Kiss_6Li,perez2023nuclear}, it is still unclear whether VQE is a scalable algorithm. Hence, the optimization procedure becomes increasingly difficult \cite{traps} with the system size because of the presence of barren plateaus in the loss landscape \cite{mcclean2018barren}. It is consequently desirable to conceive variational quantum algorithms acting on a minimal number of qubits.

Although the VQE is already a hybrid algorithm, in the sense that it relies on classical resources to perform the optimization, we take a step forward and design an algorithm relying on both neural networks and quantum circuits. More specifically, we consider VQE based on entanglement forging (EF) \cite{ibm_EF}, a circuit\hyp knitting strategy that effectively performs a Schmidt decomposition of the variational quantum state, optimizes the two sub\hyp systems separately, before reconstructing the entanglement classically. This procedure has the desirable properties of reducing the number of qubits while still reproducing the ground-state energy with high accuracy. It is similar in spirit to quantum-embedded \update{density functional theory} \cite{rossmannek2021quantum}, where quantum resources are only used for the most challenging parts. 
%The main bottleneck of this approach, being it the exponential sum over the Schmidt coefficients, can be mitigated in most cases by considering a decomposition with low entanglement between the two main sub\hyp systems. 
%In many cases, the primary bottleneck of this approach, which is the exponential sum over the Schmidt coefficients, can be alleviated by utilizing a decomposition that exhibits low entanglement between the two subsystems. Therefore, in that case, the Schmidt coefficients decay fast enough that a reasonably small cutoff can be used. Such decomposition requires a good knowledge of the target systems in general, and it can thus be difficult to choose a good set of bitstrings that will contribute to the sum.

% petite section pour dire que EF est aussi utile pour etudier l'entanglement. 
Besides computing ground state energies, EF also allows practical heuristic simulations, notably in analyzing bipartite entanglement. This concept is fundamental in quantum mechanics, as its measurement provides an understanding of the behavior of strongly correlated systems \cite{QPT_water}. For instance, bipartite entanglement has been used in condensed matter physics to study phenomena such as quantum phase transitions, topological order, and many-body localization \cite{topo_ent_entrop_kitaev, topo_order2}. Advances in experimental techniques have made it possible to measure entanglement entropy in a variety of condensed matter systems over the past few years, revealing insights into their underlying quantum properties \cite{exp_entanglement}.

The main contribution of this paper is a Schrödinger forging procedure using an autoregressive neural network (ARNN) \cite{germain2015made,van2016conditional}. This method combines the versatility of Schrödinger forging with controlling the computational resources required via the introduction of a cutoff. Generative neural networks have already been proposed for EF \cite{patrick_EF}, but only in the context of Heisenberg forging, which requires permutation symmetry of the two sub-systems. Our method, however, does not require permutation symmetry between the two sub\hyp systems, making it a more versatile approach to solving ground-state problems using quantum computers. Moreover, our algorithm naturally includes a cutoff in the number of basis states, limiting the required number of quantum circuits.

This paper is structured as follows. We first introduce EF in Sec.~\ref{Schrodinger}, as well as two ways to tackle its scalability issue based on Monte Carlo sampling and neural networks. The main contribution of this paper is then proposed in Sec.~\ref{sec:ours} as a third option.
%combining the advantages of both methods, being more accurate and 
%allowing a better control on the trade-off between expressiveness and computational expensiveness of our model. 
We conclude our work with numerical simulations in Sec.~\ref{sec:numerics} testing our hybrid architecture on various physical models, such as one-dimensional spin chains, spins on a triangular lattice with a random external field, and the nuclear shell model.

\section{Methods}
The general strategy of variational algorithms for ground state problems is to prepare a wave function ansatz $\ket{\psi}$, and using the variational principle, 
\begin{equation}
    E_0 \leq \frac{\bra{\psi}H\ket{\psi}}{\braket{\psi} },
\end{equation}
to approximate the ground state energy $E_0$ of the Hamiltonian of interest $H$. The ansatz can take the form of, e.g., a neural network \cite{Carleo_Science} or a quantum circuit \cite{vqe_original}, while the variational parameters are usually optimized with, e.g., gradient-based methods. In the following, we will explore hybrid classical\hyp quantum models aiming at describing a bipartite system with quantum circuits, while the entanglement between the partitions is forged classically.

\subsection{Entanglement Forging}
\label{Schrodinger} 
The starting point of the EF procedure is to employ a Schmidt decomposition, a direct application of a singular value decomposition (SVD), to write a quantum state $\ket{\psi}$ of a bipartite $H= H_A \otimes H_B$ quantum system, with dimensions $N_A$ and $N_B$, as

\begin{equation} \label{eq:Schmidt_decomp}
     \ket{\psi} = U \otimes V \sum_{\sigma} \lambda_\sigma \ket{\sigma}_A \ket{\sigma}_B.
\end{equation}

In the above, $U$ and $V$ are unitaries and $\ket{\sigma}_X \in \{0,1\}^{N_X}$ and $\lambda_\sigma$ are the corresponding Schmidt coefficient. The latter are positive, normalized $\sum_\sigma |\lambda_\sigma|^2 = 1$, and the number of Schmidt coefficients is called the Schmidt rank. We recall that the distribution of the Schmidt coefficients is related to the level of entanglement between the two sub\hyp systems, with the von Neumann entropy being calculated by
  \begin{equation}
     S_{vN} = -2 \sum_\sigma \lambda_\sigma^2 \log(|\lambda_\sigma|).
 \end{equation}
Therefore, maximal entanglement is characterized by a uniform distribution, while minimal entanglement by a dirac delta.

The variational state is obtained by parametrizing $U$ and $V$ with two quantum circuits and considering the Schmidt coefficients as additional variational parameters. Following \textcite{ibm_EF}, the most direct way to compute the expectation values, called \textit{Schrödinger forging}, is to directly insert the Schmidt decomposition, e.g. Eq.~(\ref{eq:Schmidt_decomp}), into $\expval{O}=\bra{\psi} \ket{O|\psi}$. Assuming that the observable $O$ admits a bipartition $O = O_A \otimes O_B$, the expectation value can then be expressed as 

\begin{equation}
\label{eq:exp_schr}
\begin{split}
    \bra{\psi}\ket{O|\psi} = & \sum_{n=1}^{2^{N/2}} \lambda_n^2 \bra{\sigma_n} \ket{U^\dagger O_AU|\sigma_n}\bra{\sigma_n} \ket{V^\dagger O_BV|\sigma_n} \\
      + &\sum_{n=1}^{2^{N/2}}  \sum_{m=1}^{n-1}\lambda_n\lambda_m  
     \sum_{p\in\mathbb{Z}_4} (-1)^p\bra{\phi_{\sigma_n,\sigma_m}^p} \ket{U^\dagger O_AU|\phi_{\sigma_n,\sigma_m}^p}  \\
    \cdot & \bra{\phi_{\sigma_n,\sigma_m}^p} \ket{V^\dagger O_BV|\phi_{\sigma_n,\sigma_m}^p},
     \end{split}
\end{equation}
where  $\ket{\phi_{\sigma_n,\sigma_m}^p}=\ket{\sigma_n}+i^p\ket{\sigma_m}$, $\mathbb{Z}_4 = \{ 0,1,2,3\}$ and all the bitstrings $\sigma$ have been labeled with a number $n$ (or $m$). Decompositions with equally sized subsystems are considered: $N_A=N_B=:N/2$. We note that this is not a strict requirement, but we \update{use} it to simplify the notations.

We remark that this involves an exponential sum in the system size. As such, two methods have been suggested to solve this scalability issue \cite{ibm_EF}. The first uses an unbiased estimator of $\bra{\psi} \ket{O_A \otimes O_B|\psi}$, that can be evaluated by importance sampling according to $\sim\lambda_n\lambda_m$. The second approach, is to leverage permutation symmetry between \update{the} two sub systems, producing  another EF scheme. Since it is defined at the operator level, we refer to it as \textit{Heisenberg forging}. This approach has been further developed by \textcite{patrick_EF} using ARNN. More details about Heisenberg forging can be found in Appendix ~\ref{an:Heisenberg_forging}.

\subsection{Schrödinger forging with generative neural networks}
\label{sec:ours}
\begin{figure*}[t!]

  \hspace{1cm}\includegraphics[width={16cm}]{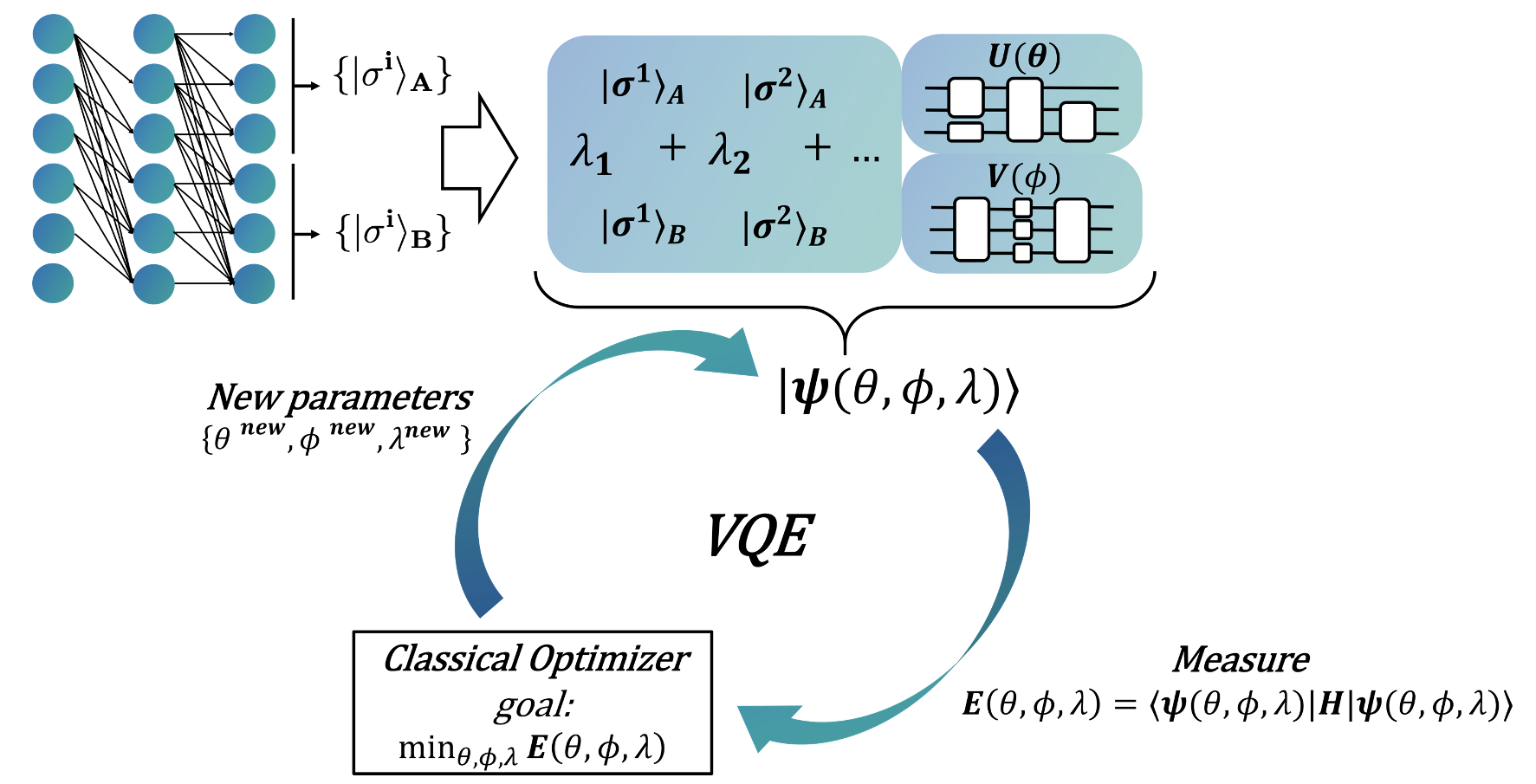}
  \caption{Schema of the end to end algorithm. The set of bitstrings is first generated by the ARNN, and is then used to perform the Schrödinger forging VQE. This involves iteratively computing the variational energy on the quantum processing unit and classically optimizing the variational parameters until convergence.}
  \label{fig:algo}
\end{figure*}

In this section, we present an approach to Schrödinger forging. The starting point is to remark that the Schmidt coefficients decay exponentially if the two subsystems are weakly entangled, as it is the case in low-energy eigenstates of chemical and spin lattice model Hamiltonians. By introducing a cutoff in the sum, it is therefore possible to improve the efficiency of the estimation while keeping a sufficiently low additive error. However, it requires a selection of a set of bitstrings among the $2^{N/2}$ total possibilities, which represents an open problem for EF. To this end, we propose to use generative models (more specifically ARNN \cite{barrett2022autoregressive}) to select the best candidates.
%, and optimizing it for the targeted application.
\update{ARNN is a type of neural network architecture commonly used in time-series forecasting and sequence modeling tasks. The autoregressive property means that the output at a given time step is regressed on its own past values. In fact, autoregressive models predict the next value in a sequence based on the previous values in that sequence.} The use of an ARNN is motivated by the fact that the Schmidt coefficients are normalized and can thus be interpreted as a probability density. Following \cite{CI}, we propose an algorithm, which is summarized in Algorithm~\ref{alg:gene_algo}. We note that this approach shares some similarities with quantum\hyp inspired genetic algorithms, see e.g., \cite{quantum_genetic}.
The parametrized unitaries and the Schmidt coefficients are finally optimized with a gradient\hyp descent\hyp based algorithm. A summary of the entire algorithm is shown in Fig.~\ref{fig:algo}.

\begin{algorithm}[H]
\caption{\textbf{Generation of the set of bitstrings}}\label{alg:gene_algo}
\begin{algorithmic}
\Inputs{Cut-off $k$}
\Outputs{Set of $k$ bitstrings}
\Initialize {Start with a random set $A$ of $k$ bitstrings}
\While{the algorithm has not converged} 
\State 1. Generate a set of bitstrings $G$ with 
\State \hspace{0.4cm} the ARNN  
\State 2. Using the bitstrings $\sigma$ in the set $A\cup G$, 
\State \hspace{0.4cm} find their Schmidt coefficient $\lambda_\sigma$ by solving the
\State \hspace{0.4cm} system of equations
\begin{equation*}
 \frac{\partial \expval{H}}{\partial \lambda_\sigma}=0 \hspace{0.5cm}  \text{with}  \hspace{0.5cm} \sum_\sigma |\lambda_\sigma|^2 = 1
\end{equation*} 
\State 3. Create the set $A'$ composed of the bitstrings 
\State \hspace{0.4cm} from $A\cup G$ with the $k$ biggest $\lambda_\sigma$
\State 4. Train the ARNN such that it models $p(\sigma) \sim |\lambda_\sigma|^2$
\State 5. Update $A \gets A'$
\EndWhile \\
\Return the set $A$ of $k$ bitstrings
\end{algorithmic}
\end{algorithm}

First, we explain how to use an auto-regressive neural network to efficiently identify the relevant bitstrings. Since the Schmidt coefficients are normalized as $\sum_\sigma |\lambda_\sigma|^2=1$, they can be interpreted as a probability density. The chain rule from probability theory can be used to write
\begin{equation}
    |\lambda_{\sigma}|^2 \sim p(\sigma_{A},\sigma_{B})  =  \prod_i p((\sigma)_i|\{(\sigma)_{j},\, j<i\}),
\end{equation}
and the bitstring pairs, associated with $\lambda_{\sigma}$ can be encoded by stacking the bitstrings of subsystem $B$ at the end of the bitstrings of subsystem $A$,
\begin{equation}
\begin{split}
    \sigma  &= \ket{\sigma_{A}, \sigma_{B}} \\
    &= \ket{(\sigma)_1,\cdots, (\sigma)_{N/2},(\sigma)_{N/2+1}\cdots, (\sigma)_N}.
    \end{split}
\end{equation}
Note that here $(\sigma)_i$ denotes the $i$th bit of the bitstring $\sigma$.

Neural networks, and more particularly, auto\hyp regressive methods are powerful tools to model such conditional densities \cite{laroche} by generating elements sequentially conditioned on the previous ones. To build the autoregressive model, we consider a dense ARNN, whose architecture is very similar to a dense feedforward neural network. The notable difference is that the weights are tridiagonal matrices, ensuring the autoregressive nature of the model. From the ARNN, the bitstrings can then be sampled directly and efficiently, as detailed in Appendix~\ref{an:sampling}. 

% \begin{comment}
% \begin{figure}[h]
%     \centering
%     \includegraphics[width={8cm}]{figures/figure_ARNN.png}
%     \caption{Auto Regressive Neural Network (ARNN) with one hidden layer used in the generative algorithm to model the probability density $p(\sigma_A^i,\sigma_B^i)$ given by the square of the Schmidt coefficients.}
%     \label{fig:ARNN}
% \end{figure}
% \end{comment}

Exploring the full space of bitstrings is exponentially difficult, motivating the use of machine learning techniques to select the basis states that contribute the most to the wave function. Inspired by the work of \textcite{CI}, we introduce an algorithm whose primary objective is to bypass exploring the extremely large space of basis states.
Starting from a random set of bitstrings $A_0$, the strategy consists of adding bitstrings generated according to the approximation of the $|\lambda_\sigma|^2$ modeled by the ARNN. Since the variational energy is quadratic with respect to the Schmidt coefficient $\lambda_\sigma$, at each iteration, they can be determined by solving the constrained linear equation system 
\begin{equation}
\begin{split}
 \frac{\partial \expval{H}}{\partial \lambda_\sigma}&=0 \\ 
 \sum_\sigma |\lambda_\sigma|^2 &= 1,
 \end{split}
 \end{equation} 
where the sum runs over the set $A\cup G$, with $A$ being the current set of bitstrings while $G$ is the set of bitstrings sampled by the ARNN. The first equation ensures that the forged wave function has minimal energy, while the second guarantees its normalization. In a second step, the current set $A$ is updated by taking the $k$ bitstrings with the highest Schmidt coefficients. The ARNN is finally trained to model $p(\sigma_A,\sigma_B) \sim |\lambda_\sigma|^2$ in a supervised way. These steps are iterated until convergence, which is reached when the current set $A$ is stable and the loss of the ARNN is close to zero. 
\update{The choice of the cutoff $k$ is usually optimized by trial and errors. Here, we start with a small cutoff, and slowly increase it until no further improvement is observed. We note that small cutoffs are often preferable, since they requires less expensive calculations, but also make the ARNN easier to train. This is why the ARNN plays an important role in choosing the optimal set of bitstrings. In summary, the training is composed of two stages: the training of the ARNN using some random initialization for $U$ and $V$, followed by the optimization of the two unitaries in order to tailor them to the set of bitstrings. The ARNN is trained to model the distribution $p(\theta)$ on the target $q$ obtained by solving the system of linear equations in Algorithm \ref{alg:gene_algo}.} 
%We note that only the bitstrings associated with the most significant coefficients are kept through the iterations and use the ARNN to directly generate the bitstrings that contribute the most to the Schmidt decomposition.
 
The choice of the loss function $\mathcal{L}$ and training set $\mathcal{T}$ play an important role in the training of the ARNN. For the training set, two possibilities are investigated: the model is either trained on the current set $A$ and the generated bistrings $G$ or, following Ref.~\cite{CI}, only on the non pruned bitstrings, i.e., the new set $A'$. Concerning the loss functions, we consider the explicit logcosh loss \cite{logcosh} and the implicit maximum mean discrepancy (MMD) loss \cite{MMD} 
\begin{equation}
    \begin{split}
    \mathcal{L}_{\mathrm{MMD}}(\boldsymbol{\theta}) & =  \sum_{\sigma_1, \sigma_2 \in \mathcal{T}} q(\sigma_1)q(\sigma_2)K(\sigma_1,\sigma_2) \\
    &  -2\sum_{\sigma_1, \sigma_2 \in \mathcal{T}} q(\sigma_1)p_{\boldsymbol{\theta}}(\sigma_2)K(\sigma_1,\sigma_2) \\ 
    & + \sum_{\sigma_1, \sigma_2 \in \mathcal{T}} p_{\boldsymbol{\theta}}(\sigma_1)p_{\boldsymbol{\theta}}(\sigma_2)K(\sigma_1,\sigma_2),
    \end{split}
\end{equation}
where
$ K(\sigma_1,\sigma_2) = e^{-\frac{||\sigma_1-\sigma_2||_2^2}{2\Delta}} $ is chosen to be a Gaussian kernel, with $||\cdot||_2$ the 2-norm and $\Delta$ the bandwidth parameter. The latter determines the width of the kernel and controls the sensitivity of the MMD measurement. A larger bandwidth allows more global comparisons, while a smaller bandwidth focuses on local details. The MMD loss function effectively minimizes the difference between the mean embedding of the two distributions. It involves a pairwise comparison of every bitstring in the training set with their contribution being controlled by the kernel.

As a benchmark, we also consider a more standard approach for modeling probability distributions using the reversed Kullback\hyp Leibler (KL) divergence for the loss of the ARNN. A detailed description of this method is presented in Appendix~\ref{ap:reversed_KL}.

%%%%%%%%%%%%%%% NUMERICS %%%%%%%%%%%%5
\section{Numerical Simulations}
\label{sec:numerics}

In this section, we present numerical experiments. 
%to illustrate the various scheme for entanglement forging.
We begin with the performance of the bitstrings selection algorithm on small models. Then, we proceed to expound upon the bitstrings selection and subsequent energy minimization process on various models of increasing complexity.

\subsection{Identify the relevant bitstrings}
\label{sec:numerics_small_models}

\begin{table*}[t!]
    \centering
    \begin{tabular}{|c|c|cccc|c|}
        \hline
        & & & & & & 
    \\[-0.8em]
         \multirow{3}{*}{Models} & standard approach &   \multicolumn{4}{c|}{\hspace{0.1cm} Proposed algorithm \hspace{0.1cm}} & \multirow{3}{*}{$\sum_{k=0}^7|\lambda_k|^2$}  \\
         & approach &   \multicolumn{3}{c}{\hspace{-0.3cm} \small{logcosh} \hspace{0.1cm}} & \hspace{0.3cm} \small{MMD} \hspace{0.1cm} &  \\
         & \footnotesize{reversed KL} &  \footnotesize{$A\cup G$} & \hspace{-0.3cm} \footnotesize{$A'$} &  \hspace{-0.3cm} \footnotesize{$A'$ and MA} & \footnotesize{$A'$} &  
    \\[-1.em]
     & & & & & & 
         \\
         \hline
        & & & & & & 
    \\[-0.8em]
        TFIM 1D 14 spins & 0/8 & 4/8 & \textbf{7/8} & \textbf{7/8} & \textbf{7/8} & 0.9641 \\
        Heis. 1D 14 spins & 0/8 & 5/8 & 5/8 & \textbf{7/8} & \textbf{7/8} & 0.9605 \\
        J1J2 1D 14 spins & 2/8 & 6/8 & \textbf{7/8} & \textbf{7/8} & \textbf{7/8} & 0.9163 \\
        TFIM 2D 12 spins & 2/8 & 6/8 & 3/8 & 6/8 & \textbf{7/8} & 0.9842 \\
        tV ($4\times3$) & 2/8 & 5/8 & 5/8 & 4/8 & \textbf{7/8} & 0.9549 
     \\[-1.em]
     & & & & & & 
        \\
        \hline
    \end{tabular}
    \caption{\textbf{Small models}: number of bitstrings generated which are among the ones with the eight biggest Schmidt coefficients in the exact decomposition. The proposed algorithm is evaluated with $A\cup G$ or $A'$ as training set $\mathcal{T}$, with or without MA. Furthermore, the ARNN if trained with the reversed KL, logcosh and MMD loss. The final column shows the sum of the truncated Schmidt coefficients.}
    \label{tab:small_model}
\end{table*}

We investigate the performance of the generative algorithm on small symmetric models: the transverse field Ising model (TFIM), the Heisenberg and $J_1$-$J_2$ model on a one\hyp dimensional (1D) chain of 14 spins with periodic boundary condition, the two\hyp dimensional (2D) TFIM on a $4\times 3$ triangular lattice with a diagonal cut and open boundary condition and the $t$-$V$ model on a $4\times 3$ grid. These models, further detailed in Appendix~\ref{an:small_models}, allow for an exact Schmidt decomposition, enabling us to assess the algorithm's performance by examining how many bitstrings associated with high Schmidt coefficients can be identified.

The results, for a cutoff dimension of $k=8$ bitstrings, are presented in Table \ref{tab:small_model}. More specifically, we can find the performance of Algorithm \ref{alg:gene_algo} in terms of the number of correctly identified bitstrings using logcosh and MMD loss. Two training sets are considered for the former: the union $\mathcal{T}=A\cup G$ and the pruned set $\mathcal{T}=A'$. Furthermore, the impact of the parameters' initialization is attenuated by model averaging (MA). The ensemble technique consists of training four ARNNs with different initial weights and taking their average as the starting point of a final ARNN. Results obtained with the more standard reversed KL approach (see Appendix~\ref{ap:reversed_KL}) are also presented. Finally, the last column of the table contains the sum of the eight highest Schmidt coefficients squared from the exact decomposition. It indicates the amount of entanglement, the cutoff's accuracy, and the probability distribution's sharpness.

The algorithm proposed in this paper is able to find the majority of the most important bitstrings. Moreover, the MMD and logcosh loss function are superior to the standard approach based on the reversed KL divergence. Indeed, with the latter, relevant bitstrings can only be found if the system is small or when the level of entanglement is high (leading to a wide probability distribution). This is not suitable in most applications, since low entanglement is important to guarantee a low additive error with a cut-off dimension. The best results are highlighted in bold, and are in general obtained with the MMD loss. More precisely, with the MMD loss, it is always able to find the four bitstrings with the highest Schmidt coefficient. This loss enables the ARNN to generalize well and make the algorithms converge quickly, as it can be further appreciated for the 2d TFIM with 12 spins in appendix \ref{an:fig_sup}, Fig.~\ref{fig:TFIM_2d_12s_histogram}. In that case, the ARNN has only seen 24 bitstrings in total during the training and it is able to find the seven bitstrings with the highest Schmidt coefficients, containing the five most important ones, in only two iterations. 

\begin{figure}[h]
    \centering
    \includegraphics[width=8cm]{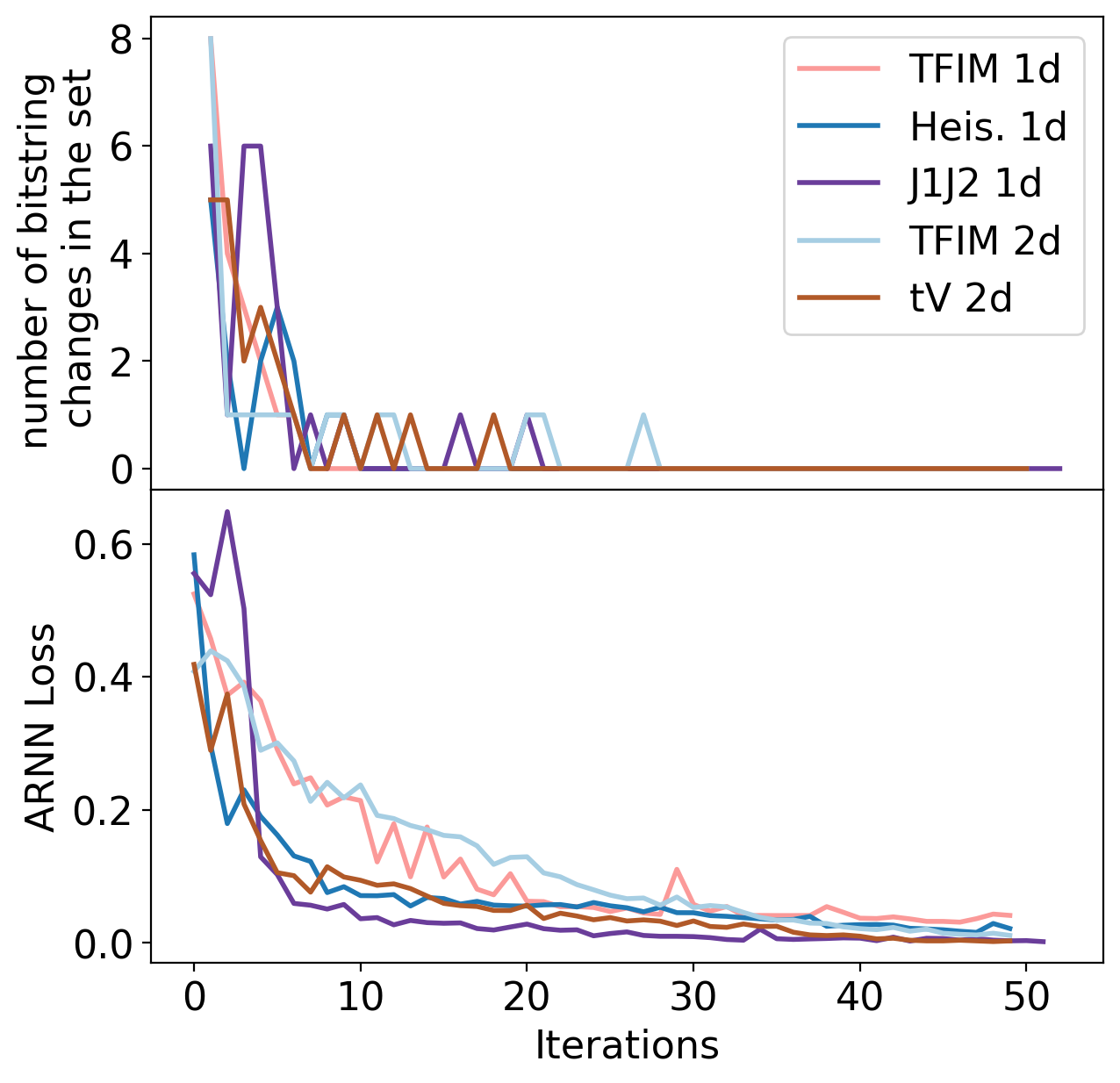}
    \caption{\textbf{Small models}. Training of the generative algorithm for different physical systems. [Top] the number of bitstrings updates between two consecutive iterations. [Bottom] value of the MMD loss at each iteration.}
    \label{fig:TFIM1d_14s_gene_algo}
\end{figure}

To gain a better understanding of the dynamics of the generative algorithm, the loss of the ARNN and the number of bitstring updates between two iterations are presented. Figure~\ref{fig:TFIM1d_14s_gene_algo} shows the results with the MMD loss on the five different Hamiltonians. In all cases, we observe that the ARNN loss converges to zero and that the generated set of bitstrings is stable.
%Moreover, the bitstring generated not present in the set of the 8 highest is the 10th highest.
%For $\mathcal{T} = A \cup G$, the ARNN encounters a greater number of random bitstrings, resulting in more oscillations. 
%In general, the dynamics can be divided into two phases: an initial phase where the loss is high and the model explores a diverse range of bitstrings, followed by a second phase where the model attempts to exploit its approximation of the probability distribution to converge and generate bitstrings with high Schmidt coefficients. The trade-off between these two phases can be modified by adjusting the number of bitstrings sampled at each iteration and the learning rate of the ARNN.

In general, the dynamics can be divided into two phases: an initial phase where the loss is high and the model explores a diverse range of bitstrings, followed by a second phase where the model attempts to exploit its approximation of the probability distribution to converge and generate bitstrings with high Schmidt coefficients. This exploration-exploitation trade-off can be modified by adjusting the number of bitstrings sampled at each iteration and the learning rate of the ARNN.

\subsection{Complete entanglement forging scheme}

In the last section, we shown that the ARNN is able to identify the bitstrings with the highest Schmidt coefficients. We now test the complete EF scheme. First, spin systems on a ring are considered, before going to a two dimensional lattice, and finally to the nuclear shell model. 

\subsubsection{Spins in one dimension}

We begin by considering the one-dimensional TFIM. More precisely, we consider a spin chain with periodic boundary conditions, an even number $N=20$ of spins, and set the coupling and the external field coefficient to one. The Hamiltonian of the model can be written as
\begin{equation}
    H = \sum_{i=1}^{N} Z^iZ^{i+1} + X^i.
\end{equation}

\begin{comment}
The resulting topology of the bipartite system is shown in Fig.~\ref{fig:chain1D}.
\begin{figure}[h]
    \centering
    \includegraphics[width={7cm}]{figures/figure_Chain_1D_PBC_sanstitre.png}
    \caption{One dimensional spin chain with Periodic Boundary conditions (PBC) and $N=20$. The blue cut represents the separation between the 2 subsystems.}
    \label{fig:chain1D}
\end{figure}
\end{comment}

\begin{figure}[h]
    \centering
    \includegraphics[width=8cm]{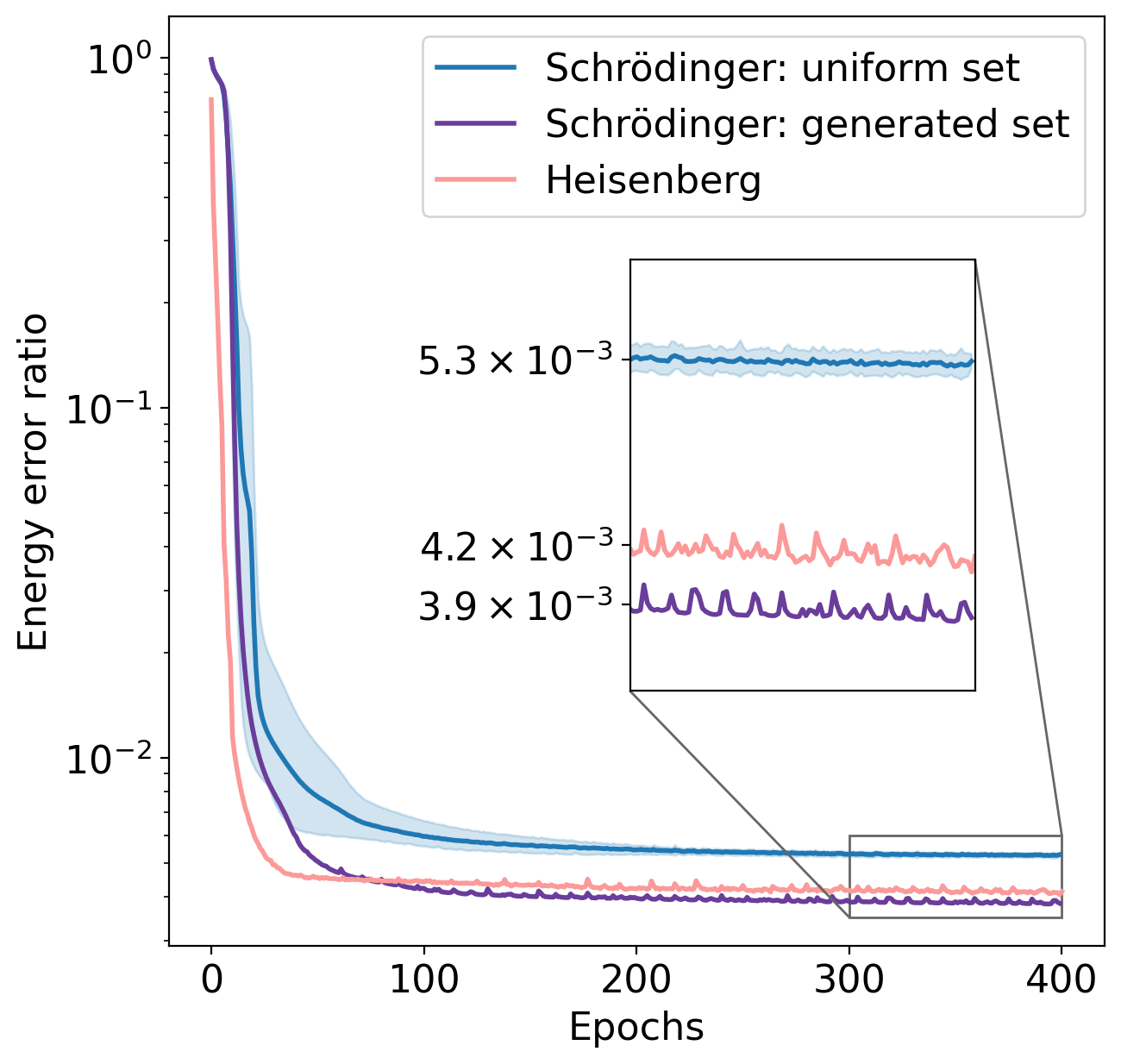}
    \caption{\textbf{1D TFIM 20 spins}. Convergence of the variational energy of forged quantum states. The blue curve represents the mean energy over ten sets of $k=8$ random bitstrings, with the shaded area displaying the standard deviation. The purple one is instead showing the training using the set generated by the ARNN. In addition, the simulation with the Heisenberg forging algorithm is shown in pink.}
    \label{fig:chain1D_k8_results}
\end{figure}

Since the system is invariant under permutation symmetry, we can compare our approach to the Heisenberg forging with ARNN. Because of the symmetry, we can choose $\sigma_A = \sigma_B$, which reduces the number of possible bitstrings. As above, a cutoff dimension of $k=8$ is chosen for the number of bitstrings, which was chosen by trial and error.
%The effect of the number of bitstrings $k$ is emphasised in Fig.~\ref{fig:chain1D_variousk}, which display the accuracy for different values of $k$. We observe that increasing $k$ has little effect on the random sampling scheme, while the ARNN quickly benefits from increasing $k$, and then plateau. The latter is probably also due to the hyperparameters during the optimization. Indeed, to be able to have a better comparison of the final results, the same hyperparameters, described in the appendix \ref{app:optimizer}, have been used. Nevertheless, to have better results, they could have been tuned for each case.
%The left panel of Fig~\ref{fig:chain1D_k8_results} shows the number of bitstrings updates in the current set, along the training loss of the ARNN, while the right panel 
Figure~\ref{fig:chain1D_k8_results} shows the energy error ratio 
\begin{equation}
    \Delta = \left|\frac{E-E_{\text{exact}}}{E_{\text{exact}}}\right|
\end{equation}
for the three forging schemes, i.e., Schrödinger with a random uniform set of bitstrings, Schrödinger with the generated set, and Heisenberg forging with the generated set. Following Ref. \cite{patrick_EF}, a pre\hyp training of the quantum circuit over 1000 iterations is performed. In both cases, the unitaries take the form of hardware efficient ansatz 
\begin{equation}
   \mathcal{U}(\Theta) = \prod_{d=0}^{D-1} \left[ U(\theta_d^0) \prod_{i=0}^{N/2}\text{CX}_{i,i+2} \cdot U(\theta_d^1) \prod_{i=1}^{N/2}\text{CX}_{i,i+2} \right] U(\theta_D),
\end{equation}
where $D=15$ is the number of layers, $\text{CX}_{i,j}$ is a CNOT gate with control qubit $i$ and target $j$, while $U(x)$ is $N$ fold tensor product of arbitrary single\hyp qubit rotation  parametrized with $3N$ parameters. We denote with $\Theta$ the set containing all indexed $\theta_i^j$. Details on the training procedure, such as values for the hyperparameters and the optimization algorithm,  can be found in Appendix~\ref{app:optimizer}.

We observe that the choice of the random set has little impact on the performance of the Schrödinger forging procedure. Moreover the models enhanced with the ARNN display better results, both being quite similar.

\begin{figure*}[t!]
    \centering
    \includegraphics[width={18cm}]{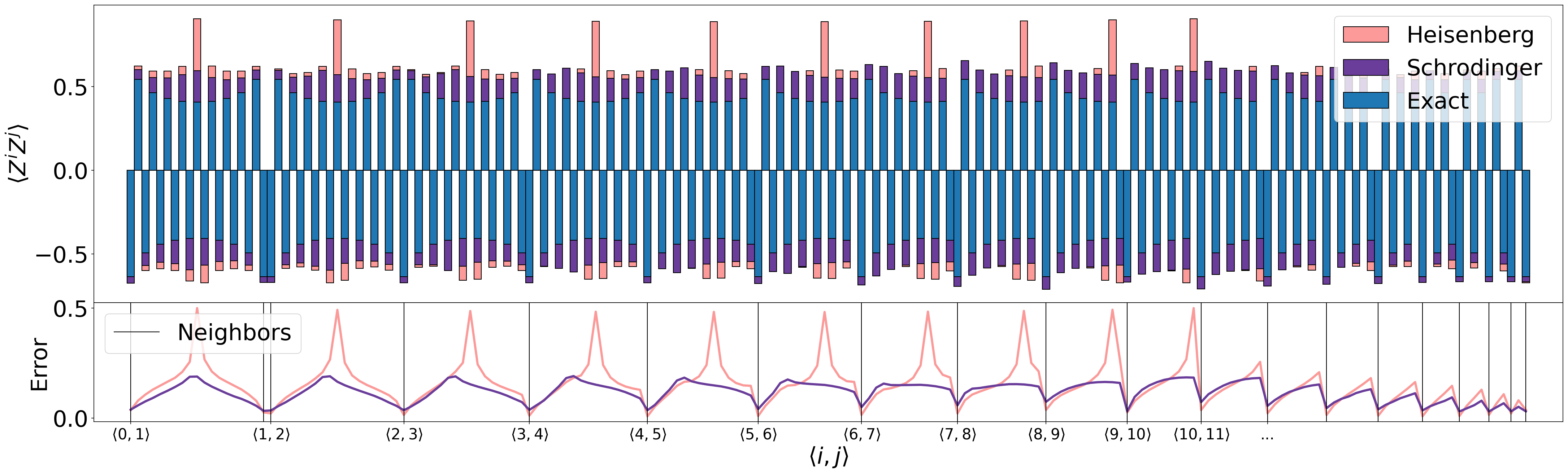}
    \caption{\textbf{Correlators in 1D}. Correlators $\expval{Z^iZ^j}$ of the Schrödinger and Heisenberg forged states on the TFIM 20 spins in 1D. The pairs $\langle i,j \rangle$ are ordered as follows: [ [$\langle i,j \rangle$ for $i<j$] for $0 \leq i <N$ ]. The neighboring cases, with $j=i+1$, are highlighted with a black vertical line. }
    \label{fig:chain1D_correlators}
\end{figure*}

To ensure that specific physical properties of the ground state, outside of its energy, are correctly reproduced, the spin\hyp spin correlators $\expval{Z^iZ^j}$ of the forged states have been calculated. They are shown in Fig.~\ref{fig:chain1D_correlators}. We observe that the accuracy is not degrading over the overlap, suggesting that the error can be explained mainly by the training of the circuits rather than the EF procedure. The error on the correlators $\expval{Z^iZ^j}$  is minimal when $j=i+1$ and maximal if the two spins are far apart in the chain. This can be explained by the locality of the ansatz, built using gates acting on neighboring qubits.

\subsubsection{Spins in two dimensions}

We now move towards two-dimensional spin lattices, which are more challenging due to local operators being mapped to non-local ones when projected onto a line. We consider the TFIM on a 2D topology described by a triangular lattice, as shown in Fig.~\ref{fig:triangular_lattices}, see Appendix~ \ref{an:small_models}. We break the permutation symmetry of the two subsystems by applying a random external field $h_i\sim U[-1,1]$. Setting the coupling constant to one, the Hamiltonian is given by
\begin{equation}
    H = \sum_{\langle i,j\rangle } Z^iZ^{j} + \sum_{i=0}^{N-1} h_iX^i,
\end{equation}
where $\langle i,j\rangle$ are neighbors according to the triangular topology. 
The triangular lattice has a high coordination number, leading to strong magnetic susceptibility \cite{blundell2001magnetism}, meaning that the system is more sensitive to external magnetic fields and can therefore exhibit stronger magnetic order and complex physical phenomena, such as, e.g., disorder, localization, and heterogeneity. 
%Hence, breaking the translation invariance induces disordered systems, while the random field causes the spins to localize in certain regions of the lattice, leading to a reduction in the effective dimensionality of the system \cite{loca_rdmh} and to spatial variations in the spin-spin interactions. This can lead to the emergence of new phases and critical behavior. 

The two-dimensional lattice is divided with a cut along the diagonal axis. We consider open boundary conditions (OBC),  cylindrical boundary conditions (CBC), and toroidal boundary conditions (TBC). Since the boundary conditions can lead to different levels of entanglement \cite{KITAEV20032, PhysRevB.82.155138}, they play an essential role in the EF procedure, which is why different configurations are considered. 

%Indeed, for OBC, the entanglement entropy follows an area law scaling,  which is a signature of the absence of long-range quantum correlations, while the entanglement entropy for TBC exhibits a logarithmic scaling, which is a signature of the presence of a critical regime \cite{PhysRevLett.90.227902}. The entanglement entropy is expected to be higher in the latter case due to interactions across the boundaries. Finally, for CBC, the entanglement entropy exhibits a hybrid scale between the logarithmic scale and the area law scale, depending on the cylinder geometry and the position of the interface. The entanglement entropy can be higher or lower than the periodic boundary condition, depending on the specific configuration of the system. This topology can be seen as a cylinder with OBCs at its extremities \cite{tri_lattice_CBC_cyl}. 

\begin{figure}[h]
    \centering
    \includegraphics[width=8cm]{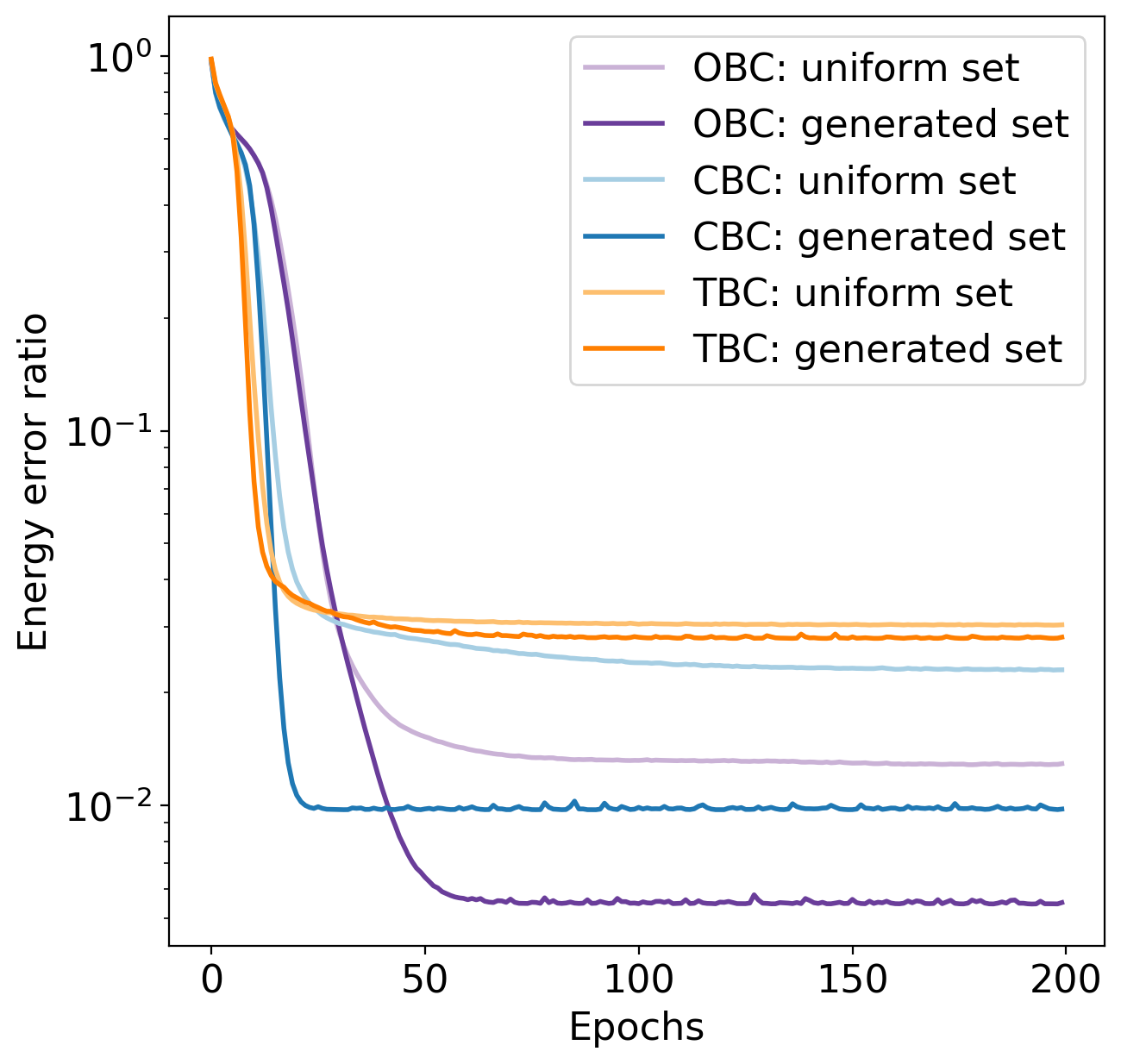}
    \caption{\textbf{2D TFIM 12 spins}. Convergence of the variational energy of forged quantum states. The colors indicate different boundary conditions, while the shaded curves show the mean energy over ten sets of k = 8 random uniform bitstrings.}
    \label{fig:energy_gene_TFIM2d12srdmh.png}
\end{figure}

%The optimization curve of the ARNN (left panel) and final energy (right panel) for the lattice with OBC and the two different cut are displayed in Fig.~\ref{fig:triangle_k8_results}. 
The convergence of the variational energies for the three boundary conditions are shown in Fig.~\ref{fig:energy_gene_TFIM2d12srdmh.png}. Like in the one-dimensional case, a cutoff of $k=8$ is chosen in the Schmidt decomposition. We observe that the bitstrings generated by the ARNN lead to an improvement of approximately $10^{-2}$ in the error energy ratio with respect to taking a random set. The most striking result, though, is that the gap between the random and generated methods is increasing with respect to the one-dimensional case, suggesting that sampling with the ARNN is becoming more effective when considering systems of increased complexity. On the other hand, no advantage can be noted in the context of TBC. It seems that the parametrization of the unitaries is the limiting factor in improving the energy error.

% could be interesting to include the spin-spin correlator of the ground state. with rdm ext field, on attend a une exp decay \cite{loca_rdmh}

\subsubsection{Nuclear shell model}

Finally, we consider light nuclei in the shell model with Cohen\hyp Kurath \cite{Hamiltonian-COHEN19651} interactions, where the Hamiltonian can be written in second quantization as
\begin{equation}
    H = \sum_i \epsilon_i \hat{a}_i^\dagger \hat{a}_i +\frac{1}{2}\sum_{ijkl}V_{ijlk} \hat{a}_i^\dagger \hat{a}_j^\dagger \hat{a}_k \hat{a}_l \ .
    \label{hamiltonian_shell}
\end{equation}
Here, $\hat{a}_i^\dagger$ and $\hat{a}_i$ are the creation and annihilation operators, respectively, for a nucleon in the state $|i\rangle$. Single-particle energies are denoted as $\epsilon_i$ and two\hyp body matrix elements as $V_{ijkl}$. The orbitals $|i\rangle = |n=0, l=1, j, j_z, t_z\rangle$, are described as functions of the radial $n$ and orbital angular momentum $l$, the total spin $j$, its projection on the $z$\hyp axis $j_z$ its projection, and the z\hyp projection of the isospin $t_z$. 

We consider nucleons in the $p$ shell model space, which includes six orbitals for the protons and six orbitals for the neutrons, while each energy is computed with respect to an inert $^4$He core. The shell-model Hamiltonian [see Eq.~\ref{hamiltonian_shell}] is converted into a qubit Hamiltonian via the Jordan\hyp Wigner~\cite{JW} transformation. Each single-particle state is represented by a qubit where $\ket{0}$ and $\ket{1}$ refer to an empty and an occupied state, respectively. Therefore, each nucleus can be distinguished by the number of excited orbitals, representing the protons and neutrons on top of the $^4$He core. 

The partition is made at the isospin level, meaning that sub\hyp system $A$ consists entirely of protons while sub\hyp system $B$ consists of neutrons. Therefore the Schr\"odinger forging is the only possible choice since the system is not symmetric under proton\hyp nucleon exchange. To build a chosen nuclei, we start from an appropriate initial state, with the desired number of nucleons, and act with an excitation preserving (EP) ansatz. EP ansätze can be built as a product of two-qubit excitation preserving blocks $U(\theta,\phi)$, also known as hop gates \cite{Panos_excitation_preserving,symmetry_economou}, of the form
\begin{equation}
    U(\theta) = \begin{pmatrix}
    1&0&0&0\\ 0&\cos{(\theta)}&\sin{(\theta)}&0 \\
    0&\sin{(\theta)}&-\cos{(\theta)}&0 \\
    0&0&0&1
    \end{pmatrix}.
\end{equation}
This set can be extended with four\hyp qubit excitation preserving gates \cite{pennylane_EP_gates}, defined as
\begin{equation}
\begin{split}
    G_{i,j,k,l}(\omega)\ket{0011} = \cos(\omega/2)\ket{0011} + \sin(\omega/2)\ket{1100} \\
    G_{i,j,k,l}(\omega)\ket{1100} = \cos(\omega/2)\ket{1100} - \sin(\omega/2)\ket{0011}.
\end{split}
\end{equation}

The parameterized circuit then takes the form of a layered ansatz composed of a product of excitation-preserving gates, where the $d$th layer is described by

\begin{equation}
\begin{split}
    \mathcal{U}(\Theta_d) &= \left(\bigotimes_{i=0}^{N-1} \text{RZ}_i(\phi_d^i)\right) \prod_{i=0}^{N/2-2}U_{2i,2i+1}(\theta_d^i)\\
    &\times  \prod_{i=1}^{N/2-3}U_{2i,2i+1}(\theta_d^i) \prod_{i=0}^{N-4}G_{i:i+3}(\omega_d^i).
    \end{split}
\end{equation}
We denote by RZ$_i(\phi$) a rotation of the $i$th qubit around the $z$\hyp axis and the subscript of the $U$ and $G$ gates indicate the qubit the gate is acting upon ($i:j$ is a slice from $i$ to $j$). The large $\Theta_d$ parameters regroup all parameters in the $d$th layer, i.e., $\Theta_d = \{\phi_d^i$, $\theta_d^i$, $\omega_d^i\}$. A sketch of the quantum circuit is depicted in Appendix~\ref{an:var_circ}.

\begin{figure}[h!]
    \centering
    \includegraphics[width={8cm}]{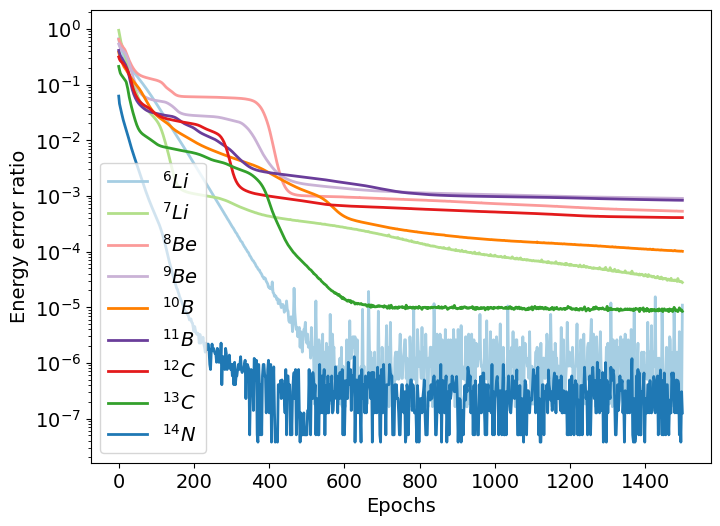}
    \caption{Convergence of the variational energy of the Schrödinger forged states corresponding to the various nuclei of the nuclear $p$ shell model. The Schmidt rank being at most 20, all bitstrings have been used in the Schmidt decomposition.}
    \label{fig:nuclear_p_shell_Econv}
\end{figure}

\begin{figure}
      \centering \includegraphics[width=7cm]{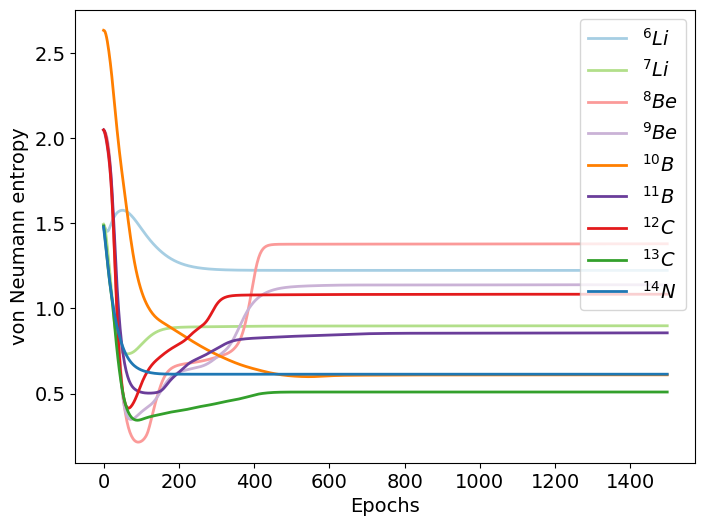}
    \caption*{(a)}
    %\hfill  
      \centering \includegraphics[width=7cm]{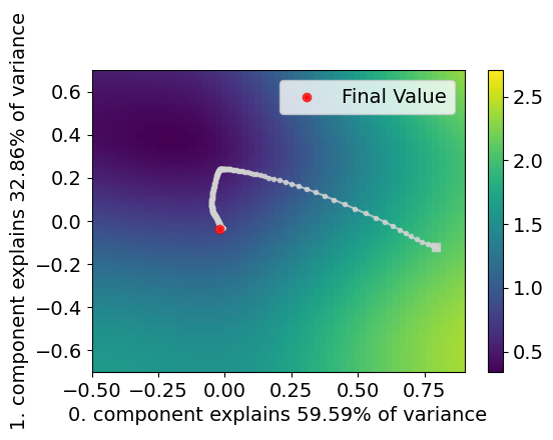}
    \caption*{(b)}
    \caption{(a) Von Neumann entropy of the various nuclei during the training. (b) Visualization of the two main components of the Von Neumann entropy of the $^{11}B$ in the variational space. In addition, the value of the entropy at each training epoch is shown (in gray) as well as the final value (in red).}
    \label{fig:p_shell_entropy}
\end{figure}

Since the Schmidt rank of the $p$ nuclear shell model is at most 20, the generative algorithm is unnecessary, as all bitstrings in the Schmidt decomposition can be used. The energy minimization for the various nuclei is presented in Fig.~\ref{fig:nuclear_p_shell_Econv}. We observe that every ground state energy in the $p$ shell can be reproduced with an error ratio of at most $10^{-3}$, even for the difficult nuclei such as $12C$. Moreover, having access to the Schmidt decomposition allows us to evaluate the von Neumann entropy, whose evolution is presented in Fig.~\ref{fig:p_shell_entropy}, which can be of broader interest. Figure~\ref{fig:p_shell_entropy}(a) shows the evolution of the von Neumann entropy during the training, while Figure~\ref{fig:p_shell_entropy}(b) displays a visualization of the von Neumann entropy in the parameter space. To this end, a principal component analysis (PCA) is performed on the entire history of the Schmidt coefficient, and a scan of the entropy along the two main components is presented. In addition, the entropy value is shown for each training epoch (in gray) and the final value (in red). 

In the final experiment, nucleons in the $sd$ shell model space, including 12 orbitals for the protons and 12 orbitals for the neutrons, are considered. For the latter, each energy is computed with respect to an inert 16O core. Using the Jordan-Wigner mapping, this model leads to a 24 qubits Hamiltonian and is composed of a total of 11'210 overlapping terms. This high number can make EF particularly expensive, as it scales linearly with it. However, since most of their coefficients are close to zeros,  an approximate Hamiltonian, consisting of the 38 overlapping terms with the most significant coefficients, is instead considered. Despite this approximation, the Hamiltonian can still reproduce 97\% of the ground state energy of the $^{23}$Na nucleus, which is the focus of this experiment.

\begin{figure}[h!]
    \centering
    \includegraphics[width={8cm}]{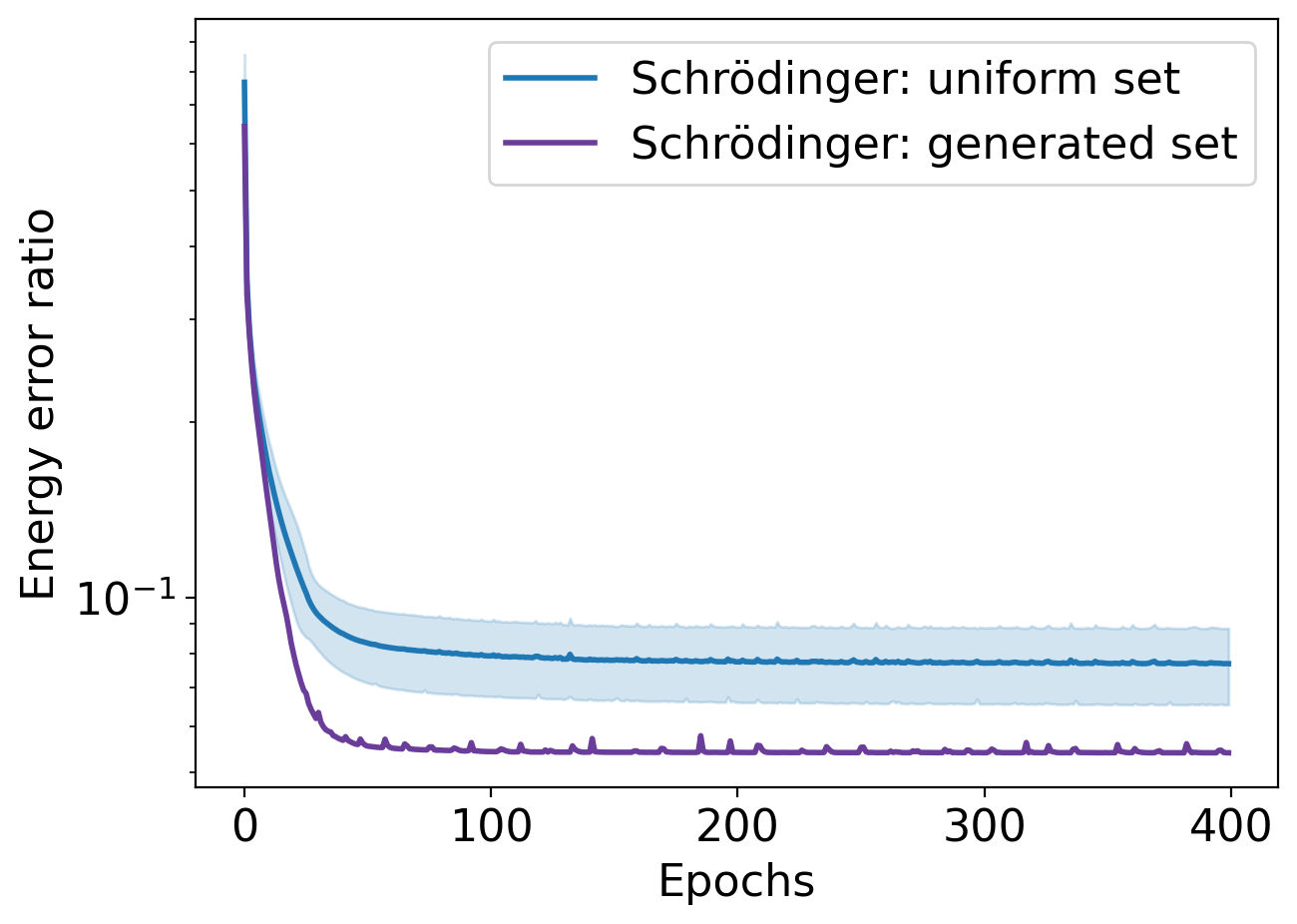}
    \caption{Convergence of the variational energy of the Schrödinger forged states corresponding to the $^{23}$Na nucleus of the nuclear $sd$ shell model. }
    \label{fig:nuclear_sp_shell_Econv}
\end{figure}

\begin{figure}[h!]
    \centering
    \includegraphics[width={8cm}]{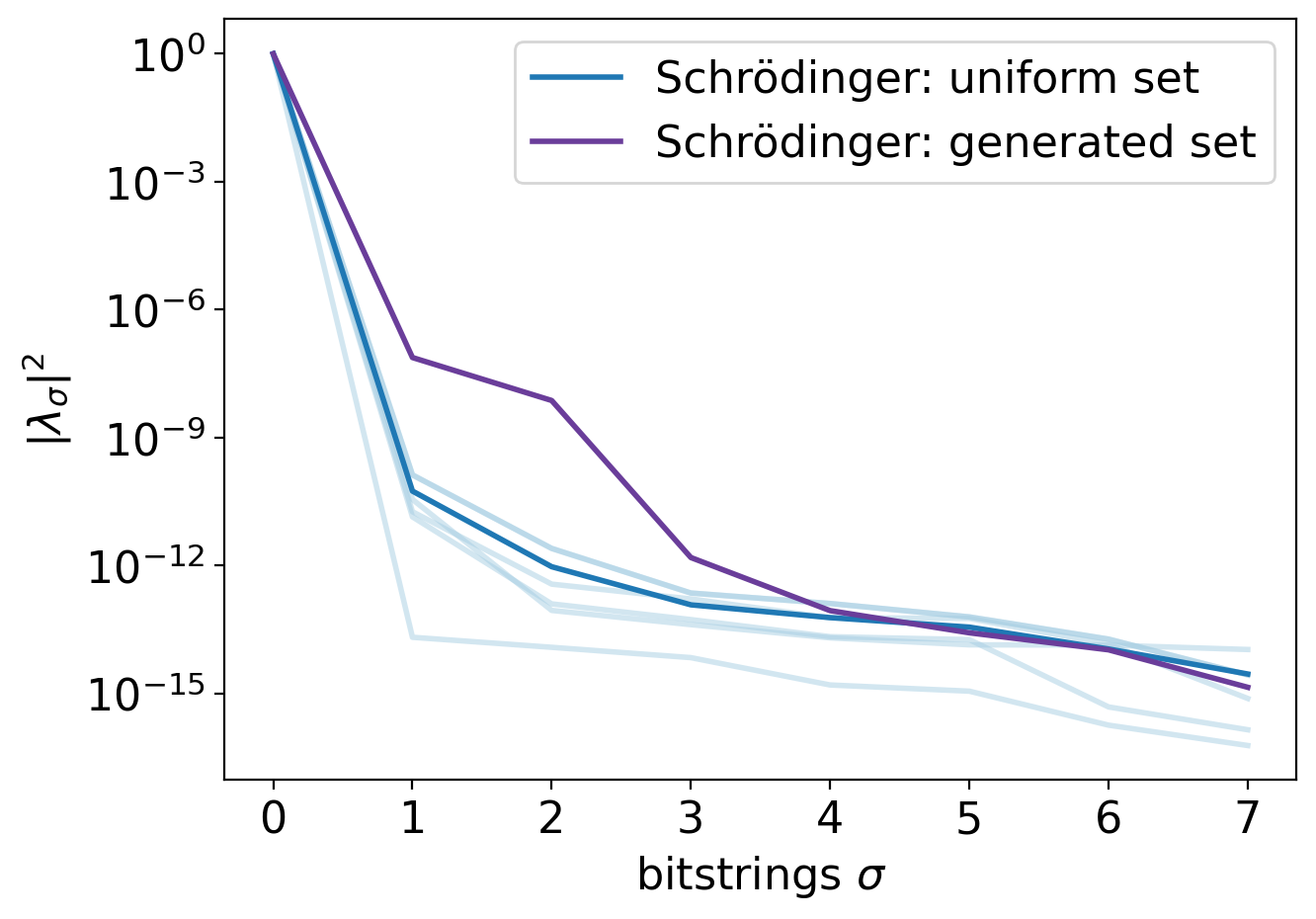}
    \caption{Final Schmidt decomposition of the variational energy of the Schrödinger forged states corresponding to the $^{23}$Na nucleus of the nuclear $sd$ shell model. }
    \label{fig:nuclear_sp_shell_Schmidt}
\end{figure}

The $^{23}$Na nucleus is composed of three protons and four neutrons on top of a $^{16}$O inert core. The ARNN sampler has therefore been modified to generate bitstrings with three ones in subsystem A (protons) and four ones in subsystem B (neutrons). The energy minimization and the final Schmidt decomposition of the $^{23}$Na are presented in Figs.~\ref{fig:nuclear_sp_shell_Econv} and \ref{fig:nuclear_sp_shell_Schmidt}, respectively. Once again, a higher accuracy is obtained with the generated set. Multiple states from the generated set are contributing to the VQE, meaning that the ARNN is useful in selecting appropriate bitstrings. On the contrary, when the random set is used, the variational circuit prefers to adapt to one state, and set the contribution from the others to zero.

\section{Discussion and Conclusion}

This paper proposes an alternative way to perform Schr\"odinger forging using autoregressive neural networks. We build on the work from \textcite{ibm_EF}, which introduced the EF-based VQE, and on~\textcite{patrick_EF}, which  efficiently compute quantum expectation values as statistical expectation values over bitstrings sampled by a generative neural network. While their work leverages the additional permutation symmetry, our work is fully general and computationally efficient due to the introduction of a cutoff dimension. Moreover, the latter is giving us additional control over the amount of quantum resources required. This is not the case in the Heisenberg forging scenario, as shown in Appendix~\ref{an:sampling}, where the ARNN begins by sampling many bitstrings and finishes by using only one. Therefore, in this specific case, the Heisenberg forging with neural networks is expensive at the beginning of the training and loses its expressive power at the end. On the other hand, Schr\"odinger forging enables better control on the trade-off between expressiveness and computational expensiveness of the variational model without having the assumption of symmetric permutation of the two subsystems. \update{When the additional permutation symmetry is present, we still recommend using Heisenberg forging, since it requires less epochs to be trained. However, we stress that many systems, such as molecules or nuclei, do not exhibit this symmetry, providing important use cases for Schrödinger forging. }

Numerical simulations have been performed on ring and triangular lattice spin systems. Schr\"odinger forging with the ARNN consistently achieves better performance for the computation of the ground state energy and correlators, compared with random sampling and Heisenberg forging with neural networks. In the case of the triangular lattice, different boundary conditions are considered, directly affecting the performance. The parametrization of unitaries is a limiting factor when complex boundary conditions are considered.
%, which the open boundary conditions seem to be the best choice. 
%When the subsystems are divided by a diagonal cut, the entanglement entropy can be influenced by the triangular geometry of the lattice. This geometry can give rise to frustration effects, which can lead to the formation of complex spin configurations that can contribute to the entanglement entropy. 

The most striking result is that the performance gap between random sampling and using the ARNN increases with the system's complexity, thus suggesting that our approach will be more profitable for larger systems. Finally, the nuclear shell model is also solved using the Schrödinger forging case up to the $10^{-3}$ error ratio for the most complex nucleus. The ARNN is unnecessary since the maximum number of possible bitstrings is 20, as all bitstrings can be used. The approach is then tested on a larger nucleus in the $sd$ shell model, $^{23}$Na. Once again, the generated set results in better accuracy than a random one.

Autoregressive models are easily interpreted and can naturally generate bitstrings with a certain number of excitations. They are also well suited for addressing the task at hand, owing to their robustness as density estimators. They do exhibit certain limitations, specifically in terms of sampling speed and the requirement for fixed-order input decomposition \cite{review_gene_model}. Nevertheless, the limited number of samples in this algorithm renders the issue of sampling speed inconsequential. Furthermore, experiments were conducted by varying the decomposition orders, and it was determined that such alterations did not yield any substantial changes in the obtained results. Masked multilayer perceptrons are a straightforward choice for building the autoregressive model. However, other architectures could be more suitable in some cases. In particular, transformers \cite{transformer} provide a strong alternative since they are highly parallelizable and efficient in capturing the global context and long-range dependencies due to their attention mechanism. 

At the beginning of this work, simulations were carried out on small models. In these cases, all bitstrings could be taken into account during the VQE. It was observed that the order of the bitstrings, with respect to their coefficient (in absolute value), did not significantly change during the VQE. Therefore, choosing the bitstrings at the beginning of the circuit training enables us to perform well. However, it could be suitable in some cases to train the quantum circuits and ARNN simultaneously, taking advantage of parameter sharing.

Finally, we note that there is not necessarily a correlation between having sets of bitstrings associated with high Schmidt coefficients and the trainability of the corresponding variational state. Indeed, in some cases, taking a set of bitstrings with lower Schmidt coefficients might be favorable to make the variational circuits easier to train. 
%Therefore, we could choose the bitstrings which maximize the gradients, or adapt the form of each variational circuit,  an approach close to the ADAPT-VQE \cite{ADAPT-VQE}.
Therefore, it may be possible to include this feature in the algorithm by choosing bitstrings that maximize the gradients. Alternatively, an algorithm, adapting the form of each variational circuit, an approach close to the ADAPT-VQE \cite{ADAPT-VQE}, could be investigated.

%les gain performances sont pas enorme mais c'est dure d'isoler le facteur limitant. Lors des simul, le cutoff+choix de set est surement pas le seul facteur limitant, beaucoup d'autre points peuvent limiter les perf (ansatz des circuits, optimizers, optim strat...).

\section*{Code availability} 
The numerical simulations of the quantum circuits have been performed with Pennylane \cite{pennylane}, powered by a JAX backend \cite{jax}, while the $\mathrm{NETKET}$ library \cite{netket} has been used for the ARNN. Solving the constraint systems of equations in the generative algorithm involves a projected gradient descent algorithm available in the JAXopt library \cite{jaxopt}. Moreover, the Heisenberg forging code is available on Github \cite{patrick_git}. Visualization of the evolution of the von Neumann entropy is performed using orqviz \cite{orqviz}.
The Python code of this project is accessible on Github \cite{de_Schoulepnikoff_Learning_Schmidt_decompositions}.

\section*{Acknowledgement}
The authors thank A. Mandarino and T. Papenbrock for stimulating discussions as well as for the computation of the nuclear shell model's matrix elements. O.K., M.G and SV are supported by CERN through the CERN Quantum Technology Initiative. 

%\bibliography{bibliography}
%\bibliographystyle{plain}
\bibliography{bibliography_li6} 
% \bibliography{bibliography}

\begin{comment}

\end{comment}

\newpage

\onecolumngrid

\appendix

%\section{Schrödinger forging: Expectation values} 
%\label{an:exp_val}

\section{Heisenberg Forging}
\label{an:Heisenberg_forging}
In the section, we briefly cover the basis of Heisenberg forging, covered in more details in Refs.~\cite{ibm_EF,patrick_EF}.
In this scenario, we assume a symmetric bipartition, i.e., with $U_A = V_B \equiv U$ and find a more efficient way to compute the expectation value. We first need to decompose $O$ as
\begin{equation}
    O_A\otimes O_B + O_B\otimes O_A = \frac{a_0}{2}\left(\{O_A,O_B\}\otimes \mathbb{1} + \mathbb{1}\otimes \{O_A,O_B\} \right) +\sum_{\alpha,\beta \in \{0,1\}}a_{\alpha,\beta}C_{\alpha,\beta}^*\otimes C_{\alpha,\beta},
\end{equation}
where $\{\cdot,\cdot\}$ denotes the anti\hyp commutator, $|a_{\alpha,\beta}|\leq 1$ are real coefficients, and $C_{\alpha,\beta}$ $n$\hyp qubit Clifford operators that are defined below. Combining this with the Schmidt decomposition, and by symmetrizing the observable, we obtain  
\begin{equation}
\label{eq:heisenberg}
    \langle \psi|O|\psi\rangle = a_0\sum_n \lambda_{\sigma_n}^2\Re(\langle \sigma_n|U^\dagger O_A O_B U|\sigma_n \rangle) +\sum_{\alpha,\beta \in \{0,1\}}\frac{a_{\alpha,\beta}}{2} \sum_{n,m}\lambda_{\sigma_n} \lambda_{\sigma_m}|\langle \sigma_m|U^\dagger C_{\alpha,\beta} U|\sigma_n\rangle|^2.
\end{equation}
Since $O_A,\, O_B \in \{\mathbb{1},X,Y,Z\}^{\otimes N/2}$, we either have $[O_A,O_B]=0$ or $\{O_A,O_B\}=0$. In the first case, we can find a Clifford circuit $V$ such that $O_A = VZ_pV^\dagger$ and $O_B=VZ_qV^\dagger$. We can then define 
\begin{equation}
    C_{\alpha,\beta} = \frac{1}{2}V\left(\mathbb{1}+(-1)^\alpha Z_p + (-1)^\beta Z_q - (-1)^{\alpha+\beta}Z_pZ_q \right)V^\dagger
\end{equation}
In the remaining case, we can simply use $C_{0,0} = (O_A+O_B)/\sqrt{2}$, $C_{0,1} = (O_A-O_B)/\sqrt{2}$ and $a_{1,0}=a_{1,1}=0$.

The estimation of the sums can be performed non-trivially, using Monte Carlo sampling, where the number of samples grows as $1/\epsilon^2$, with $\epsilon$ the additive error. However, the sampling step is not obviously scalable in the Schrödinger case, and it will be discussed below. We also point out that, despite a sampling overhead, the individual quantum circuits are easier to implement than without the Schmidt decomposition because they are shallower and require fewer qubits.

\section{Sampling from the ARNN} 
\label{an:sampling}
\begin{figure}[h!]
    \centering
    \includegraphics[width={7cm}]{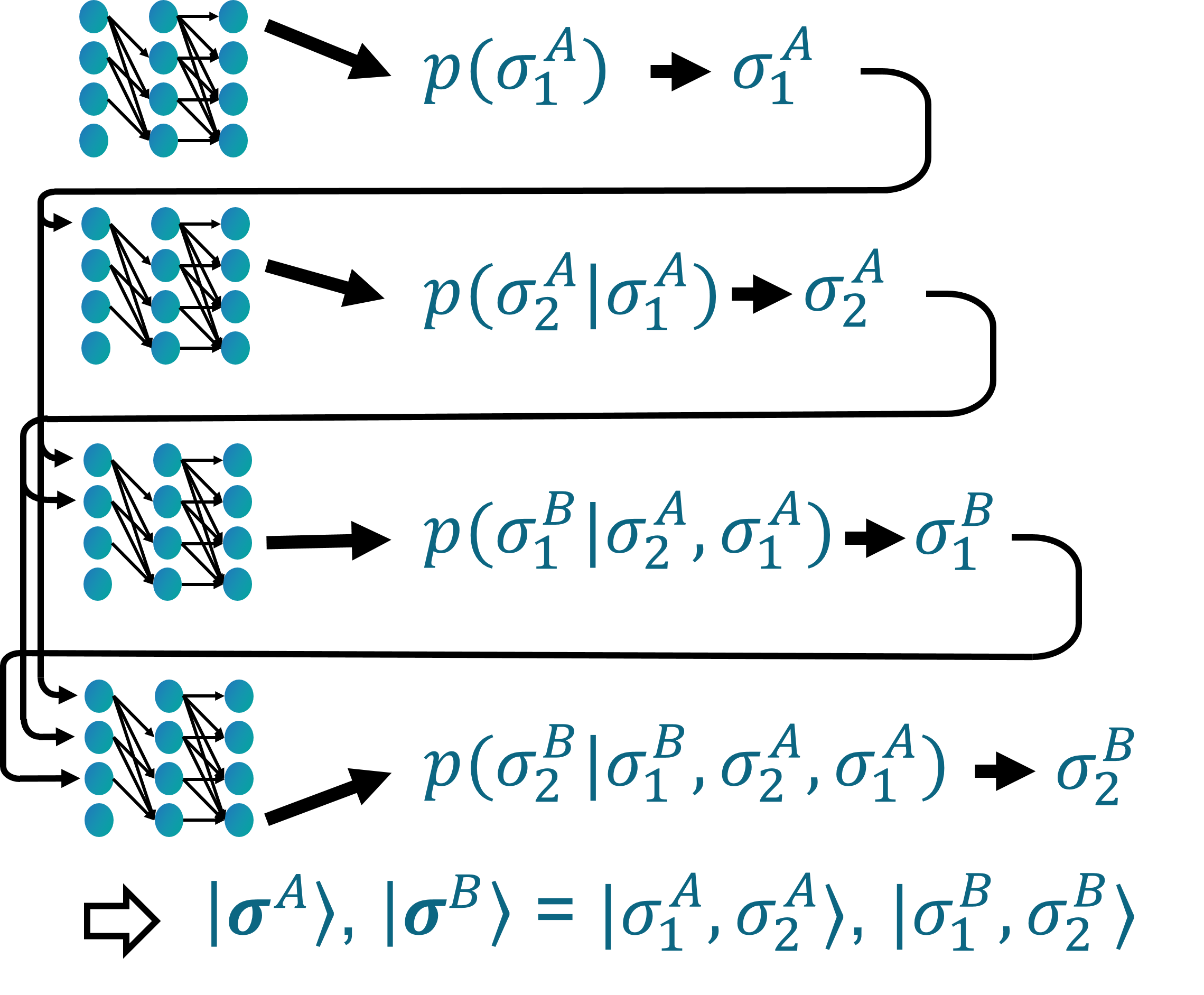}
    \caption{Description of the sampling procedure of a bitstring with the ARNN. Each bit of the bitstring is generated sequentially and randomly, according to the probability modeled by the network. The latter is the output of the ARNN with the inputs being the values of the previous bit, already generated. The illustrated situation corresponds to a system of four qubits divided equally into two sub-systems of two qubits. \update{Note that the picture describes in fact one ARNN used in parallel to sample the different bits, and not multiples ARNN used in a sequential manner, as could be induced by the arrows in the diagram.}}
    \label{fig:sampling}
\end{figure}

In this section, we provide more details on how to efficiently and directly sample the bitstrings with the ARNN. We proceed recursively, as shown in Fig.~\ref{fig:sampling}. We begin by sampling the first bit of the string, which is then given as an input to sample the second bit, and so on up to the last bit.

In the nuclear shell model, it is important to control the number of value\hyp one bit appearing in the string, since each nucleus is defined by a certain amount $k$ of excited orbitals. Thus, the same procedure can be slightly modified to generate bitstrings with a fixed number of ones. Indeed, we just need to change the conditional probability in the sampling procedure, which can be done by setting $p(\sigma_i|\{(\sigma)_j,\, j<i\}) = 0$ if $\sum_{j<i} (\sigma)_j = k$. This ensures a maximum of $k$ excitation. If, on the other hand, there is only $l<k$ excitation at the end of the string, the last $k-l$ bits are turned into one to correct for it. While this leads to non-uniform sampling at the beginning of the training, we expect the ARNN to overcome this issue by incorporating it through the learning stage.

 \begin{figure}[h!]
     \centering
     \includegraphics[width={7cm}]{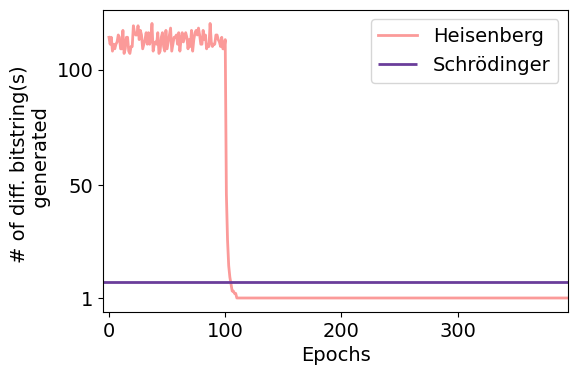}
     \caption{Comparison between the number of different bitstrings sampled for Heisenberg and Schrödinger forging. In the Heisenberg case, at the beginning, a large number of states are required to be prepared on the quantum computer, while at the end, only one state remains. For Schrödinger forging instead, we have full control over the number of states we want to prepare.}
     \label{fig:nbr_bitstrings_Heis_vs_Schr}
 \end{figure}

A notable difference between the Schr\"odinger and Heisenberg forging schemes is that for the latter, it is impossible to control how many states one has to prepare on the quantum hardware. Indeed, in this case, all bitstrings sampled by the ARNN must be taken into account. In practice, as shown in Fig.~\ref{fig:nbr_bitstrings_Heis_vs_Schr}, many states must be prepared at the beginning of the training and only one at the end. In the case of Schr\"odinger forging, since the cut-off can be fixed at the beginning, the number of states to be prepared on the hardware is constant. Following Ref.~\cite{patrick_EF} a 1000-epoch pre training on the unitaries has been performed as proposed in Ref.~\cite{patrick_EF}. However, other optimization strategies could be considered.

% \section{Supplementary Figures} 
% \label{an:fig_sup}

% In this section, we present several figures. Being less essential or taking a lot of space, these last ones have not been put in the main text. We present them anyway in the appendix because, for some interested readers, they can be useful.

% \begin{figure}[h!]
%         \begin{minipage}[b]{0.48\linewidth}
%       \centering \includegraphics[width=\textwidth]{figures/entropy_convervence_sns_sanstitre.png}
%     \caption*{(a)}
%     \end{minipage}
%     \hfill
%     \begin{minipage}[b]{0.48\linewidth}   
%       \centering \includegraphics[width=\textwidth]{figures/entropy_trajectory_B11_sanstitre.png}
%     \caption*{(b)}
%     \end{minipage}
%     \caption{(a) Von Neumann entropy of the various nuclei during the training. (b) Vizualization of the Von Neumann entropy of the $^{11}B$ in the variational space. To do this, a PCA was performed on the entire history of the Schmidt coefficient. Then, a scan on the two principal components was carried out. In addition, the value of the entropy at each training epoch is shown (in grey) as well as the final value (in red). This was done using orqviz \cite{orqviz}. }
%     \label{fig:p_shell_entropy}
% \end{figure}

\section{Overview of the many-body Hamiltonians of the small models}
\label{an:small_models}

Here, we present the many-body quantum Hamiltonians used for the numerical simulations. First, we consider spins models: the TFIM, Heisenberg and $J_1$-$J_2$ model on a 1d chain and the 2D TFIM on a triangular lattice. We also consider fermionic models, such as the $t$-$V$ model on a 4$\times$3 grid and the nuclear shell model. 
%The presentation is divided into two parts: first the models being symmetric under the permutation of the two subsystems and then, the models without permutation symmetry of the two subsystems.
%The topologies of the spin models 1d and 2d are presented in Figure \ref{fig:chain1D_14s} and \ref{fig:triangular_lattice_12s}, respectively. 

\begin{figure}[h!]
        \begin{minipage}[t]{0.48\linewidth}
      \centering \includegraphics[width=6cm]{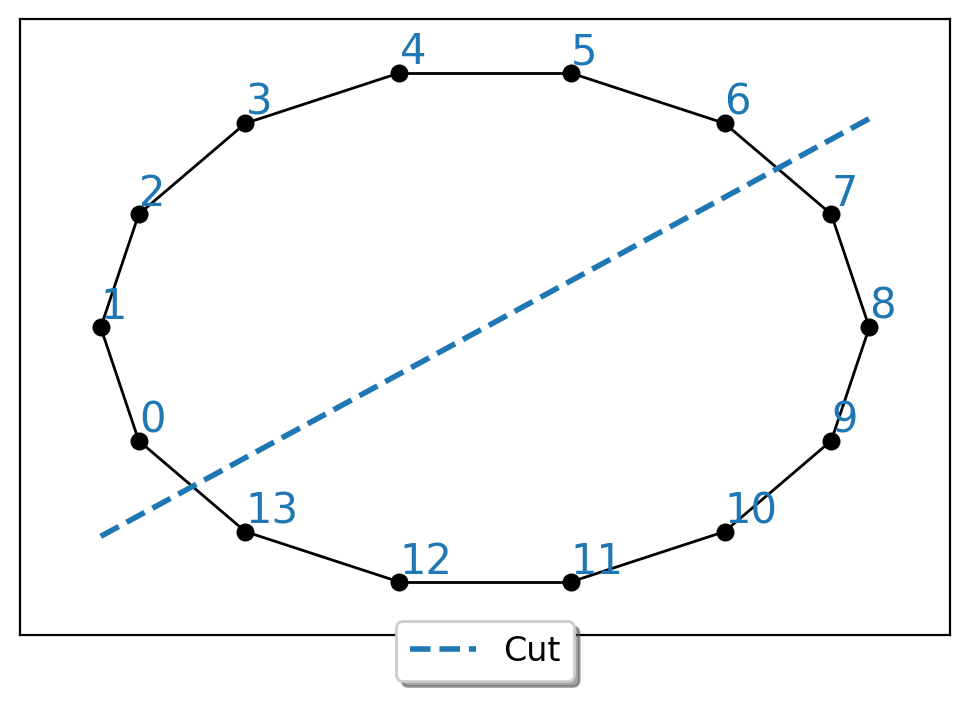}
    \caption*{(a)}
    \end{minipage}
    \hfill
    \begin{minipage}[t]{0.48\linewidth}   
      \centering \includegraphics[width=5.4cm]{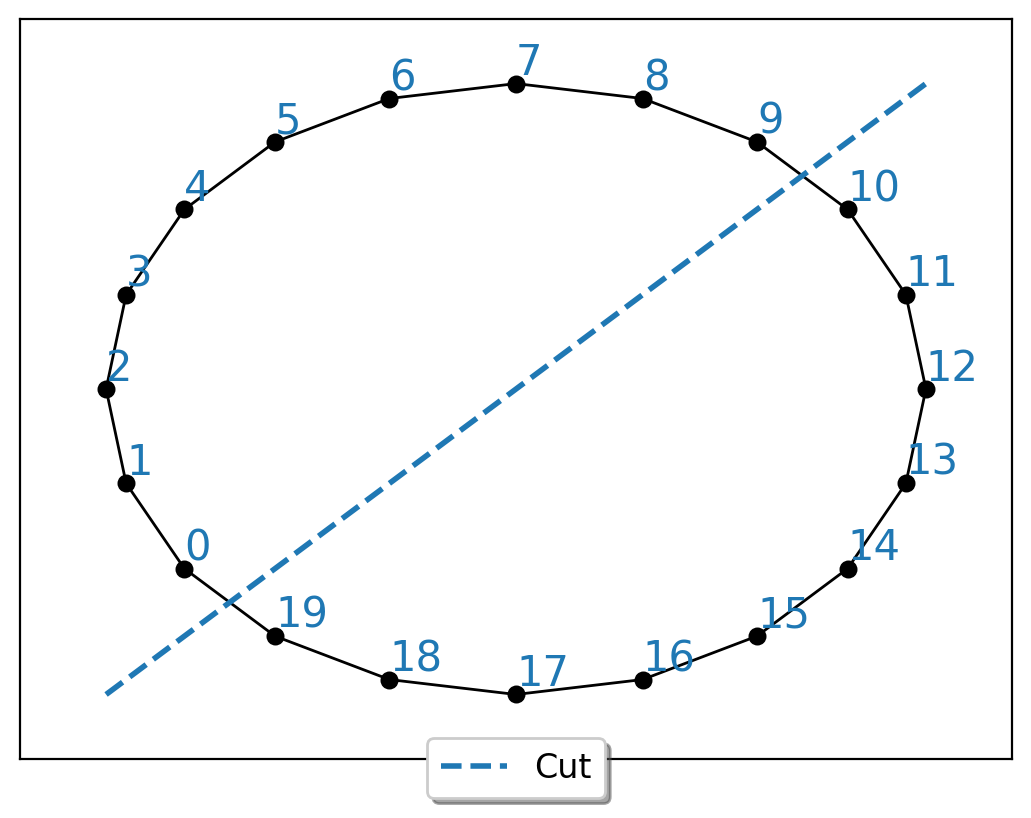}
    \caption*{(b)}
    \end{minipage}
    \caption{One dimensional spin chain with PBC, $N=14$ spins (a) and $N=20$ spins (b). The blue cut represents the separation between the 2 subsystems.} 
    \label{fig:chain1D_14s_20s}
\end{figure}

%\vspace{0.4cm}
%\noindent \textit{Permutation Symmetric Models} %\vspace{0.1cm}\\
The Hamiltonians of the 1D TFIM is 
\begin{equation} \label{eq:TFIM1d}
    H = J \sum_{i=0}^{N} Z^iZ^{i+1} + X^i.
\end{equation}
The Hamitonian of the 1D Heisenberg model is 
\begin{equation}
    H = J \sum_{i=0}^N X^iX^{i+1} + Y^iY^{i+1} + Z^iZ^{i+1},
\end{equation}
while for the 1D $J_1$-$J_2$ model 1d we have
\begin{equation}
    H = J_1 \sum_{i=0}^N X^iX^{i+1} + Y^iY^{i+1} + Z^iZ^{i+1} + J_2 \sum_{i=0}^N X^iX^{i+2} + Y^iY^{i+2} + Z^iZ^{i+2}.
\end{equation}
For these models, $J=1$, $J_1=1$, $J_2=0.2$, and periodic boundary condition (PBC), i.e., $ N\equiv0$, $N+1\equiv1$, are used. The topology of the spin chain with the separation between the two subsystems is presented in Fig.~\ref{fig:chain1D_14s_20s} (a) and (b) for 14 and 20 spins, respectively.  \\

\begin{figure}[h!]
        \begin{minipage}[t]{0.48\linewidth}
      \centering \includegraphics[width=6cm]{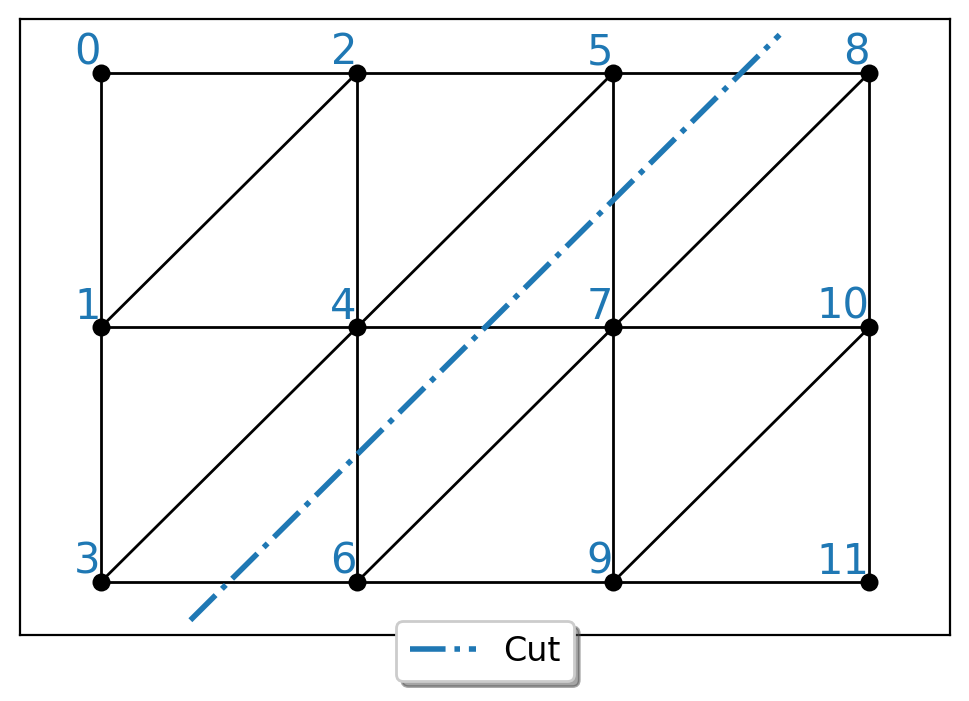}
    \caption*{(a)}
    \end{minipage}
    \hfill
    \begin{minipage}[t]{0.48\linewidth}   
      \centering \includegraphics[width=6cm]{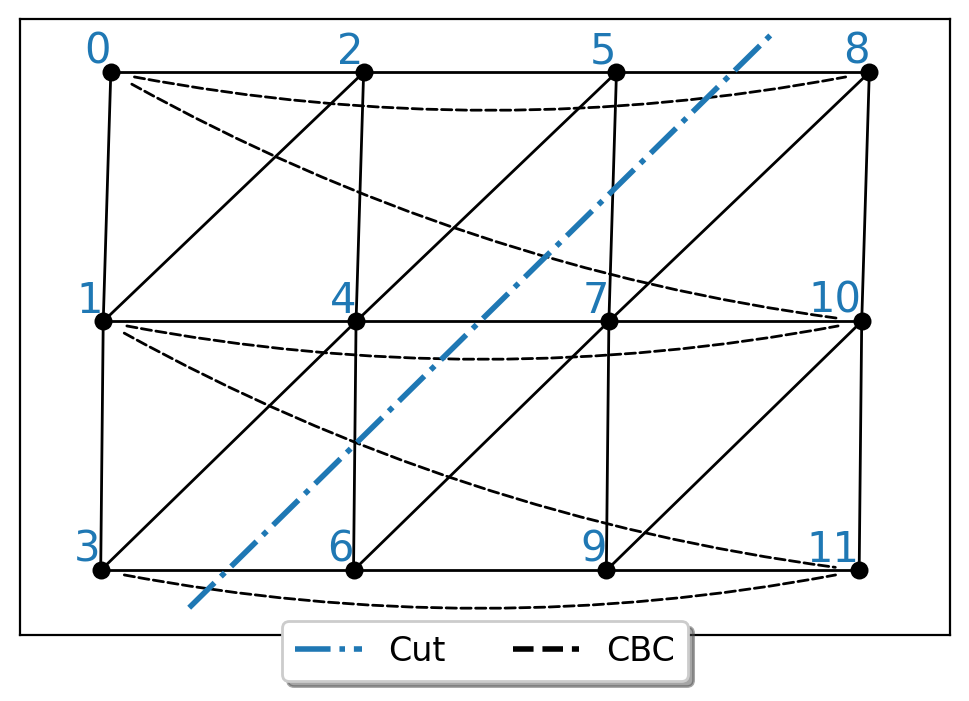}
    \caption*{(b)}
    \end{minipage}
    \begin{minipage}[t]{0.48\linewidth}
      \centering \includegraphics[width=6cm]{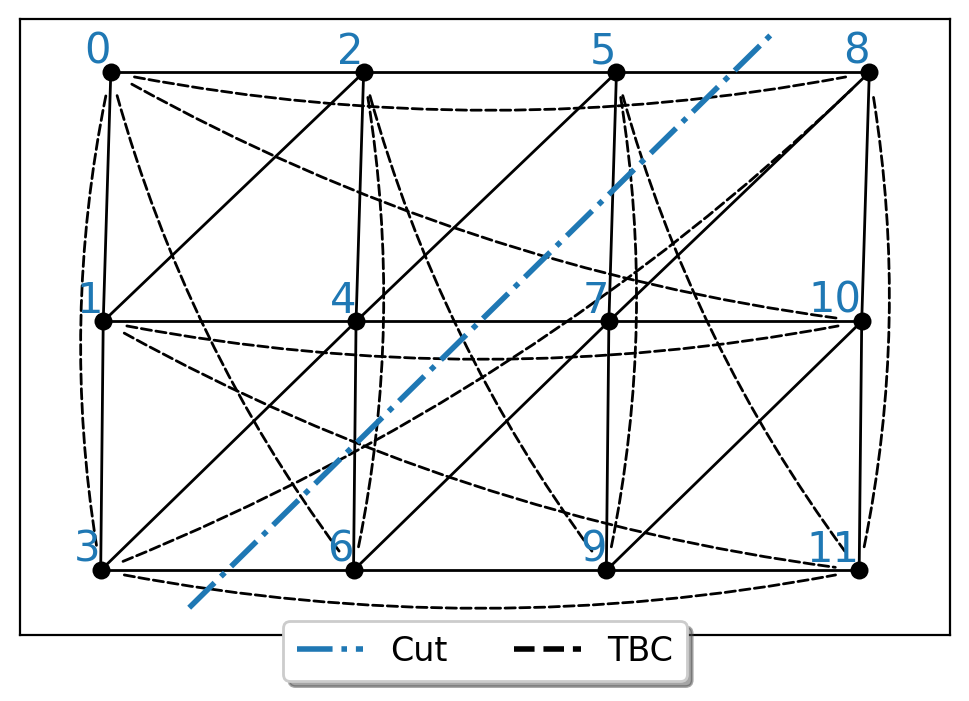}
    \caption*{(c)}
    \end{minipage}
    \caption{Triangular lattices used for the simulations. Lattices of 12 spins with OBC, CBC and TBC are shown in (a), (b) and (c) respectively. The two subsystems are defined with a diagonal cut (blue). } 
    \label{fig:triangular_lattices}
\end{figure}

\noindent The Hamiltonian of the 2D TFIM is given by
\begin{equation}
    H = \sum_{\langle i,j\rangle } Z^iZ^{j} + \sum_{i=0}^{N-1} X^i,
\end{equation}
where $\langle i,j\rangle$ are neighbors according to the triangular topology, see Fig.~\ref{fig:triangular_lattices}, which also shows the different cuts and boundary conditions. This model is more challenging due to local operators being mapped to non-local ones when projected onto a line. Moreover, it has a high coordination number which leads to a strong magnetic susceptibility \cite{blundell2001magnetism}, meaning that the system is more sensitive to external magnetic fields and can exhibit stronger magnetic order. 

The Hamiltonian of the $t$-$V$ model is 
\begin{equation}
    H = -t\sum_{\langle i,j\rangle } (a_i^\dagger a_j + a_j^\dagger a_i) + V\sum_{\langle i,j\rangle}a_i^\dagger a_i a_j^\dagger a_j,
\end{equation}
with $a_i$ and $a_i^\dagger$ being respectively the creation and annihilation operators on site $i$. A $4\times3$ system of spinless fermions with periodic boundaries and $t=V=1$ is considered. It is mapped to a qubit Hamiltonian with the Jordan-Wigner transformation. In this model, fermions are allowed to  move on the grid, modifying the energy of the system. In this spinless version, there is only one spin-orbit per site, giving a final Hamiltonian of 12 qubits.

\section{Modelling the probability distribution with a more standard approach} \label{ap:reversed_KL}

In this section, we present a more standard approach for modeling probability distributions using the reversed KL divergence for the loss of ARNN. We show why it is unsuitable for the considered problem situation.

This approach aims to model the full probability distribution $|\lambda_\sigma|^2$. Since we only have samples from the approximated probability distribution $p(\sigma_A,\sigma_B)$, the reversed KL can be used to learn the best representation of the distribution self-consistently. Thus, at each iteration, the training set comprises bitstrings sampled from the approximation distribution given by the ARNN. The latter is then trained in a supervised way to model the target distribution $|\lambda_\sigma|^2$ by minimizing the reversed KL divergence 
\begin{equation}
   \mathcal{L}_{\text{ARNN}}^{\text{rev-KLD}} =  \mathop{\mathbb{E}}_{\sigma \sim p} \Big[ \log\frac{p(\sigma_A,\sigma_B)}{\lambda_\sigma^2} \Big].
\end{equation}
With this choice, wherever $p(\sigma_A,\sigma_B)$ has a high probability, $\lambda_\sigma^2$ will also take a high value. This mode-seeking behavior is desired since the objective is mainly to sample bitstrings associated with a high Schmidt coefficient. 

 \begin{figure}[h!]
     \centering
     \includegraphics[width={7cm}]{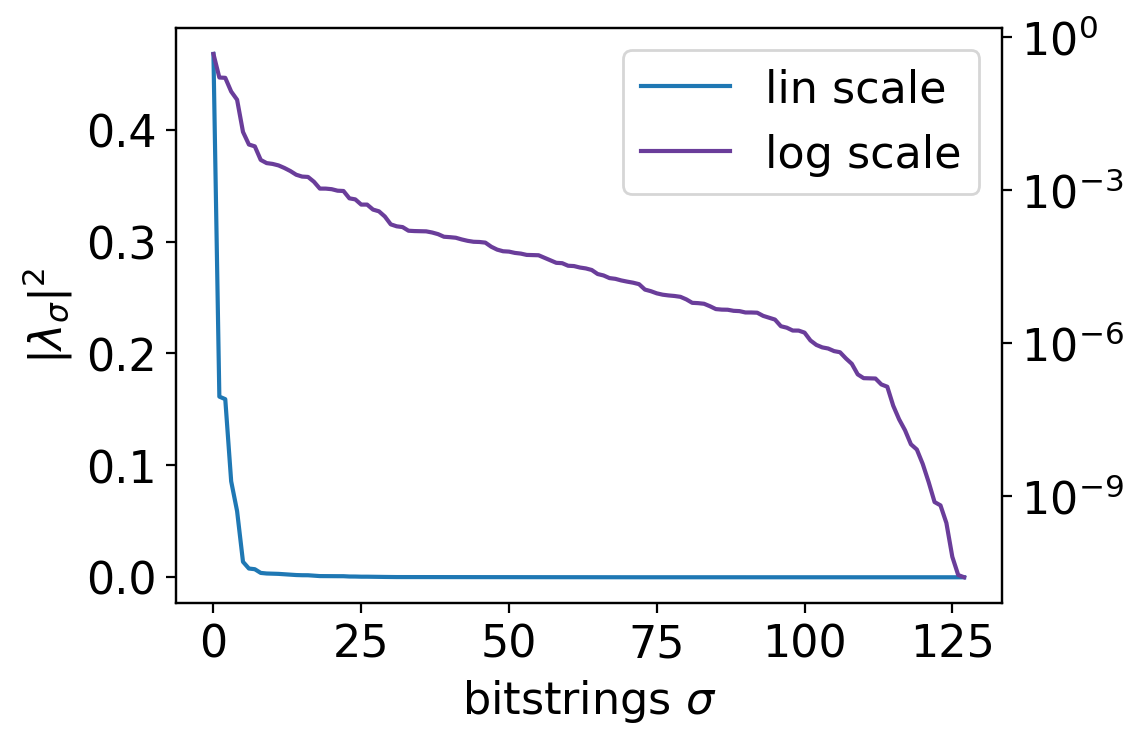}
     \caption{Exact Schmidt decomposition on the TFIM 14 spins with periodic boundary conditions. The coefficients have been arranged in descending order and are presented with a linear (blue) and a logarithmic (purple) scale.}
     \label{fig:schmidt_exact_decomp_TFIM14s}
 \end{figure}

With weakly entangled systems, which is desirable to have a low additive error with the cut-off in the Schmidt decomposition, the target probability distribution is very sharp, as shown in Fig.~\ref{fig:schmidt_exact_decomp_TFIM14s}. Such probability densities are very difficult to model with this approach. Indeed, with high probability, the training sets are composed of bitstrings associated with very small Schmidt coefficients. In this flat region (left of Fig.~\ref{fig:schmidt_exact_decomp_TFIM14s}), the probability density appears uniform, and it is challenging to extrapolate the relevant bitstrings. Moreover, due to the normalization constraint, the Schmidt coefficients of a small set of bitstrings are not good estimators of the Schmidt coefficients of the ground truth distribution.

However, this defeats our purpose of identifying bitstrings with a high Schmidt coefficient rather than modeling the entire probability distribution. Hence, adopting a training strategy that keeps the bitstrings with high Schmidt coefficients through the iterations is convenient. With such a training strategy, employing a loss composed of an average over the model data samples is impossible. The explicit form of the reversed KL divergence 
\begin{equation}
   \mathcal{L}_{\text{ARNN}}^{\text{expl-rev-KLD}} =  \sum_\sigma  p(\sigma_A,\sigma_B) \Big[ \log\frac{p(\sigma_A,\sigma_B)}{\lambda_\sigma^2} \Big],
\end{equation}
would be an alternative if only the target probability distribution is not very sharp. Hence, the reversed KL divergence is not a symmetric measure. Consequently,  the gradients obtained from the reversed KL divergence may not provide stable and robust updates for the model when the predicted distribution diverges significantly from the target distribution. This lack of robustness makes it challenging to learn in highly uncertain situations or when the model needs to adapt to changes in the training set, which is the case here. The logcosh and MMD loss were therefore used since they are more robust and suitable for modeling sharp distributions. Indeed, they do not suffer the same limitations since they focus on individual samples rather than the overall distribution and do not overemphasize outliers.

\section{Optimization details and hyperparameters}
\label{app:optimizer} 

In this section, details on the optimization procedure are given. During the VQE stage of the training, the adabelief optimizer \cite{adabelief} is used to update the quantum circuit parameters, while Nesterov's accelerated gradient descent scheme \cite{nesterov} is performed for the Schmidts coefficient. One iteration of the Schmidts coefficient is done every ten iterations of the circuit parameters. The hyperparameters of the adabelief optimizers, following the convention of the original paper, are $\beta_1=0.9$, $\beta_2=0.999$ and $\epsilon=10^{-16}$, while for Nesterov, a momentum coefficient of 0.6 is used. In both cases, we set the learning rate between 0.1 and 0.01. 

The generative algorithm is trained using adabelief with the same hyperparameters and a learning rate of 0.001. The ARNN comprises five hidden layers and a hidden neuron density of $\alpha=2$. At each iteration of the generative algorithm, the ARNN samples between 10 and 50 bitstrings to build the set $G$, while the exact value has been manually tuned for each simulation. This influences the performance since low values cause the ARNN to converge very quickly, leading to spikes caused by the lack of generalization and overfitting. On the other hand, high values deteriorate the algorithm's computational efficiency, convergence speed, and memory requirement in the same way as batches in stochastic gradient descent. For ARNN, the Lecun normal initializer for the weight, zero initial biases, and scaled exponential linear unit (SELU) activation function $\lambda = 1.0507$ $\alpha = 1.6733$ were used \cite{selu_activation}.

 \section{Variational Circuits} 
 \label{an:var_circ}
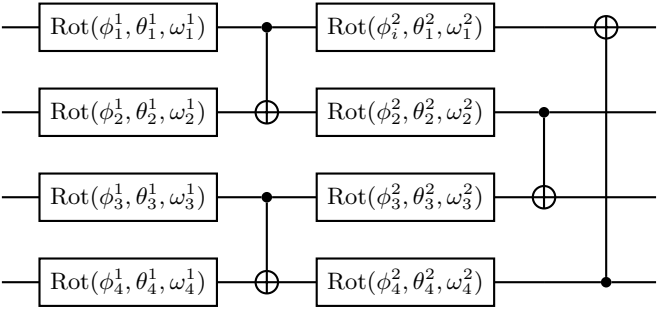
\begin{figure}[h!]
     \centering
     \begin{quantikz}
     & \gate{\mathrm{Rot}(\phi_1^1,\theta_1^1,\omega_1^1)} & \ctrl{1} & \gate{\mathrm{Rot}(\phi_i^2,\theta_1^2,\omega_1^2)} & \qw & \targ{} & \qw   \\
     & \gate{\mathrm{Rot}(\phi_2^1,\theta_2^1,\omega_2^1)} & \targ{} & \gate{\mathrm{Rot}(\phi_2^2,\theta_2^2,\omega_2^2)} & \ctrl{1}& \qw & \qw  \\
     & \gate{\mathrm{Rot}(\phi_3^1,\theta_3^1,\omega_3^1)} & \ctrl{1} & \gate{\mathrm{Rot}(\phi_3^2,\theta_3^2,\omega_3^2)} & \targ{} & \qw & \qw  \\
     & \gate{\mathrm{Rot}(\phi_4^1,\theta_4^1,\omega_4^1)} & \targ{} & \gate{\mathrm{Rot}(\phi_4^2,\theta_4^2,\omega_4^2)} & \qw & \ctrl{-3} & \qw  \\
     \end{quantikz}
     \caption{Variational quantum circuit used for parametrizing the unitaries in the Schmidt decomposition for spin and fermionic models.}
     \label{fig:TFIM_circuit_layer}
 \end{figure}

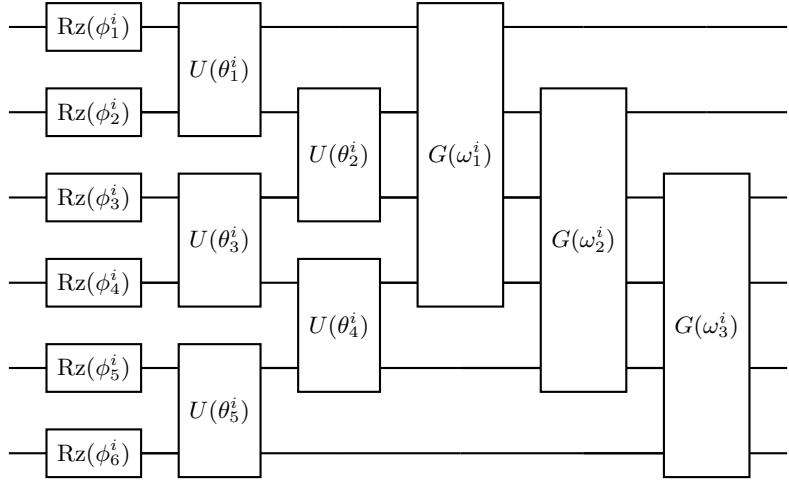
\begin{figure}[h!]
 \label{fig:circuit-xanadu}
     \centering
     \begin{quantikz}
     & \gate{\mathrm{Rz}(\phi_1^i)} & \gate[wires=2][0cm]{U(\theta_1^i)} &  \qw & \gate[wires=4][0cm]{G(\omega_1^i)} & \qw & \qw & \qw  \\
     & \gate{\mathrm{Rz}(\phi_2^i)} & \qw & \gate[wires=2][0cm]{U(\theta_2^i)} & \qw & \gate[wires=4][0cm]{G(\omega_2^i)} & \qw & \qw   \\
     & \gate{\mathrm{Rz}(\phi_3^i)} & \gate[wires=2][0cm]{U(\theta_3^i)} & \qw & \qw & \qw & \gate[wires=4][0cm]{G(\omega_3^i)} & \qw   \\
     & \gate{\mathrm{Rz}(\phi_4^i)}&  \qw & \gate[wires=2][0cm]{U(\theta_4^i)}  & \qw & \qw & \qw & \qw  \\
     & \gate{\mathrm{Rz}(\phi_5^i)}&  \gate[wires=2][0cm]{U(\theta_5^i)}  & \qw & \qw & \qw & \qw & \qw  \\
     & \gate{\mathrm{Rz}(\phi_6^i)}&  \qw & \qw  & \qw & \qw & \qw & \qw  \\
     \end{quantikz}
    
      \caption{Variational quantum circuit used for parametrizing the unitaries in the Schmidt decomposition for the shell models. The gates correspond to the ones described in the main text.}
      \label{fig:SHELL_circuit_layer}
 \end{figure}

 In this section, we provide a visual example of the quantum circuits used for VQE ansätze. A layer of the variational circuit used for the spin systems is shown in Fig.~\ref{fig:TFIM_circuit_layer} for $N=4$ qubits, while Fig.~\ref{fig:SHELL_circuit_layer} shows one layer of the circuits used for the nuclear shell model. We use the notation Rot to describe a generic rotation around the Bloch sphere, composed of $Ry$, $Rz$ and $Ry$ rotation. 

 \newpage

\section{Evolution of the generated set} \label{an:fig_sup}

\begin{figure}[h]
    \centering
    \includegraphics[width=8cm]{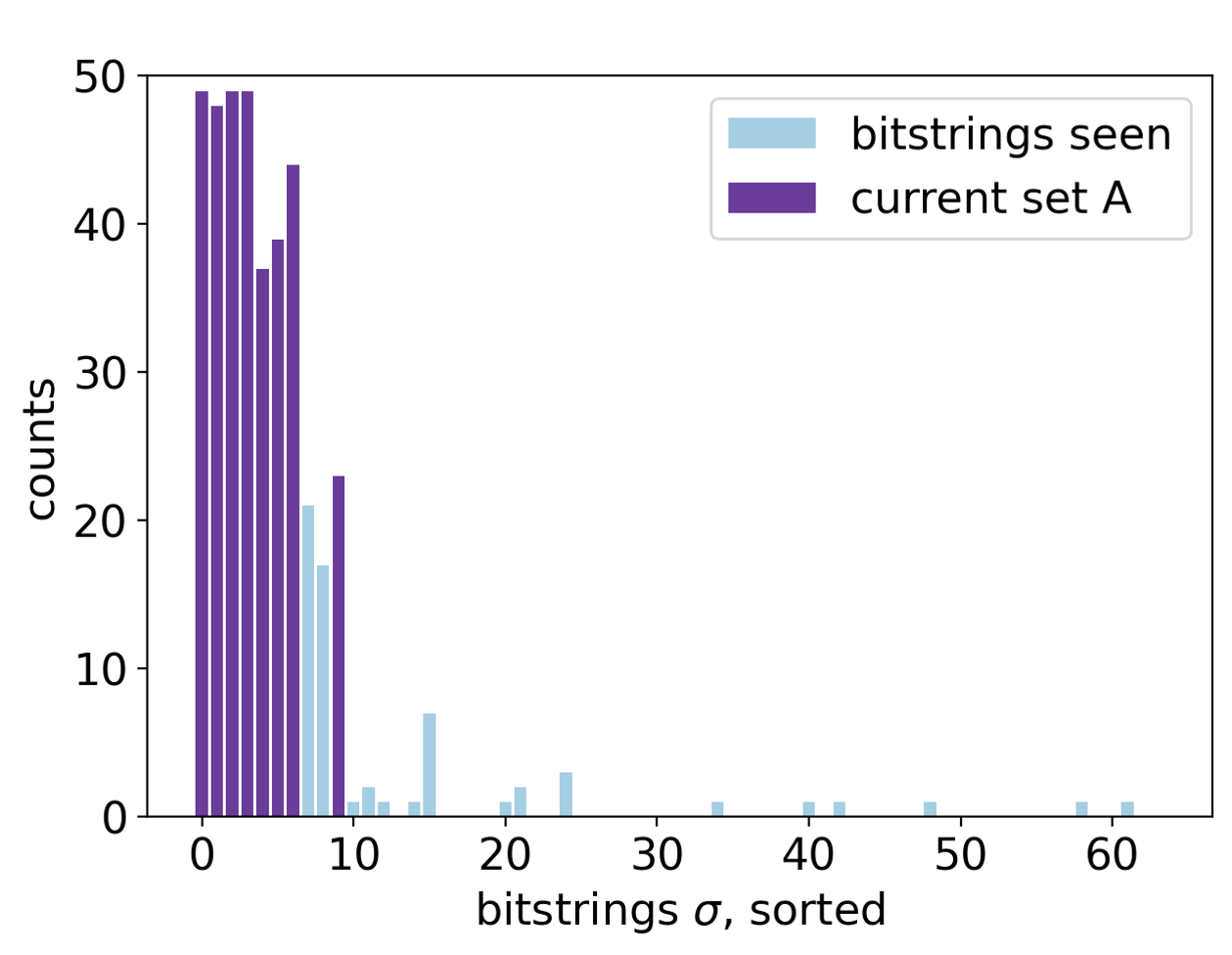}
    \caption{\textbf{2D TFIM 12 spins, MMD loss}: Histogram showing the bitstrings seen by the ARNN (present in the set $A'$) with the respective number of times (counts). The results after 50 iterations are presented. The bitstrings contained in the final set $A'$ are highlighted in purple and the bitstrings seen previously are illustrated in light blue. }
    \label{fig:TFIM_2d_12s_histogram}
\end{figure}

A histogram was produced to visualize the ARNN training set evolution. Shown in Fig.~\ref{fig:TFIM_2d_12s_histogram}, it illustrates the number of times each bitstring has been present in the training set $\mathcal{T}=A'$ of the ARNN. The bitstrings present in the final set are shown in purple and those present during the algorithm are shown in light blue. The illustrated example is at the end of the algorithm on the 2D TFIM 12 spins with the MMD loss. Bitstrings are ordered in such a way that their associated Schmidt coefficient decreases (in absolute value). The bitstrings associated with the highest Schmidt coefficient are the most frequently viewed by the ARNN.

At the end, the seven bitstrings associated with the biggest Schmidt coefficient are present in the final set. The eighth bitstring in the set is bitstrings number ten. Given that bitstrings eight and nine were seen during the algorithm and that their associated Schmidt coefficients squared are very low and close to the one of bitstring ten, we can explain this lack by numerical errors when determining the Schmidt coefficients.

\end{document}